\documentclass[twocolumn,prd,nofootinbib,floats,floatfix,amsmath,amssymb,secnumarabic]{revtex4} %
\pdfoutput=1
\usepackage{amsmath}
\usepackage{amsfonts}
\usepackage{amssymb}
\usepackage{latexsym}
\usepackage{graphicx}
\usepackage[english]{babel}
\usepackage{multirow}
\usepackage{float}
\usepackage{url}
\usepackage{slashed}
\usepackage{verbatim}
\usepackage{hyperref}

\usepackage{ulem,fancyvrb}
\usepackage{xcolor} 

\newcommand{\zc}[1]{{\color{black}{#1}}}
\newcommand{\jc}[1]{{\color{black}{#1}}}

\newcommand{\be}{\begin{equation}}
\newcommand{\ee}{\end{equation}}
\newcommand{\ba}{\begin{array}}
\newcommand{\ea}{\end{array}}
\newcommand{\bea}{\begin{eqnarray}}
\newcommand{\eea}{\end{eqnarray}}

\renewcommand{\d}{\mathrm{d}}
\def\roughly#1{\mathrel{\raise.3ex\hbox
{$#1$\kern-.75em\lower1ex\hbox{$\sim$}}}}

\def\po{$\phantom{1}$}

\newcommand{\e}[1]{e^{#1}}
\newcommand{\de}{{\mathbf e} }
\newcommand{\pd}{{\mathbf p} }
\newcommand{\dH}{{\mathbf H} }

\begin{document}

\title{An optimistic CoGeNT analysis}
\author{James M. Cline\footnote{jcline@physics.mcgill.ca}}
\author{Zuowei Liu\footnote{zuoweiliu@physics.mcgill.ca}}
\author{Wei Xue\footnote{xuewei@hep.physics.mcgill.ca}}
\affiliation{Department of Physics, McGill University,
3600 Rue University, Montr\'eal, Qu\'ebec, Canada H3A 2T8}

\begin{abstract}    Inspired by a recently proposed model of
millicharged atomic dark matter (MADM), we analyze several classes of
light dark matter models with respect to CoGeNT modulated and
unmodulated data, and constraints from CDMS, XENON10 and XENON100. 
{After removing the surface contaminated events from the original
CoGeNT data set,} we find an acceptable fit to all these data (but
with the modulating part of the signal making a statistically
small contribution), using somewhat relaxed assumptions about
the  response of the null experiments at low recoil energies, and
postulating an unknown modulating background in the CoGeNT data
at recoil energies above 1.5 keVee. We
compare the fits of MADM---an example of inelastic magnetic dark
matter---to those of standard elastically  and inelastically
scattering light WIMPs (eDM and iDM).  The iDM model gives the best
fit, with MADM close behind. The dark matter interpretation of the
DAMA annual modulation cannot be made compatible with these results
however.  We find that the inclusion of a tidal debris component 
in the dark matter phase space distribution improves the fits or
helps to relieve tension with XENON constraints.

\end{abstract}
\maketitle

\section{Introduction}

It is an exciting time, albeit puzzling, for direct detection of dark
matter.  The three experiments DAMA \cite{Bernabei:2008yi}, CoGeNT 
\cite{Aalseth:2010vx,Aalseth:2011wp} and CRESST \cite{Angloher:2011uu}
see hints of light DM of mass $\sim 10$ GeV and cross section $\sim
10^{-40}$ cm$^2$ for scattering on nucleons.  Although much effort
has been made to determine whether DAMA and CoGeNT best-fit regions
(and possibly those of CRESST) can be compatible 
\cite{Fitzpatrick:2010em}-\cite{Foot:2012rk}, it is difficult to 
find any dark matter model which 
achieves this quantitatively, and which moreover is consistent with 
the null results of  XENON10 \cite{Angle:2011th},
XENON100  \cite{Aprile:2011hi}, CDMS
\cite{Ahmed:2009zw,Ahmed:2010wy,Ahmed:2012vq}, and most recently EDELWEISS
\cite{edelweiss}.\footnote{Ref.\ \cite{edelweiss} appeared as we were
finishing this work, and is therefore not included in our analysis. 
Since it obtains weaker limits than XENON, we do not expect this to
affect our conclusions.}\ \ It seems unlikely that all of the
hints of positive detections can be correct since even nonstandard DM
models with many adjustable parameters give a poor fit to the global
data \cite{Frandsen:2011ts,Schwetz:2011xm,Farina:2011pw}.  Arguably,
the best hope for reconciling the positive and negative signals
centers around the CoGeNT results, which can be compatible with the
lightest and most weakly coupled DM models, hence those having the
greatest chance of being marginally compatible with the null results. 
This is our motivation for focusing the present study on CoGeNT.  

Another of the difficulties with CoGeNT is that it presents an annual
modulation amplitude that seems to be too large compared to the
unmodulated rate unless a nonstandard velocity distribution is invoked
for dark matter in the galactic halo \cite{Fox:2011px,McCabe:2011sr,Arina:2011zh}.  This problem
is only exacerbated by recent results indicating that the earlier
published events included contamination from surface events 
\cite{collar}.  If CoGeNT is indeed seeing
dark matter, it is tempting to speculate that there is some
contamination of the modulating part of the signal that should be
subtracted, to ameliorate this tension.  We explore the effect of
including such a background, finding that the CoGeNT-allowed regions
of parameter space are hardly affected by the modulating part of
the data, regardless of any such background, due to their lower 
statistical significance relative to the time-averaged part of the 
signal.

Although we also consider standard elastic and inelastic dark matter
in the present study, 
one of our motivations for carrying it out was the 
proposal of a new class of millicharged atomic dark matter  (MADM)
models in \cite{Cline:2012is} (a simpler variant of models previously
discussed in refs.\ \cite{Kaplan:2009de,Kaplan:2011yj}), with emphasis on the special case where
the dark ``electron'' and ``proton'' constituents have equal masses,
$m_\de = m_\pd$.  In the ref.\ \cite{Cline:2012is}, rough estimates were made as to
the parameter values required to explain the CoGeNT observations. 
Here we better quantify those by direct comparison to the data.  
We will show that MADM gives an acceptable fit to the data for the dark
$\dH$ atom mass {$m_\dH \sim 10$ GeV, the  hyperfine mass splitting
$\delta m_\dH \sim 25$ keV, and the fractional charge of the $\de$ and
$\pd$ constituents $\epsilon\sim 0.01$,}  
which controls the scattering cross section.  
{The fit to the CoGeNT data by MADM is better than the one of 
standard elastic dark matter (eDM), and only slightly worse
than that of standard inelastic dark matter (iDM), which
requires similar values for the mass and mass splitting. 
}

It is interesting to note that in the special $m_\de = m_\pd$ case of
interest, the effective charge of dark atoms is zero even at short
wavelengths, due to the cancellation between the charge clouds of the
two constituents, and the model becomes an example of magnetic
inelastic dark matter \cite{Chang:2010en}-\cite{Weiner:2012cb} (this identification has not
been previously realized).  Magnetically interacting dark matter has
recently been promoted as a model that can fit ``everything,'' all the
null and positive direct detection constraints \cite{DelNobile:2012tx},
including DAMA and CRESST.    In the present work, we will not reach
such optimistic conclusions,
 even though we find some room for CoGeNT
to be compatible with the Xenon and CDMS constraints, {if we exercise
a moderate degree of skepticism concerning the claimed sensitivity of
the Xenon experiments to low recoil energy events.}\footnote{We are not willing to consider
$7\sigma$-$8\sigma$ allowed regions of DAMA for the purpose of finding 
overlap, as done in \cite{DelNobile:2012tx}}

We also consider the effect of astrophysical uncertainties,  including
nonMaxwellian velocity distributions  \cite{Lisanti:2010qx} and
contributions from  debris flow \cite{Lisanti:2011as,Kuhlen:2012fz}. 
We find quantitative changes to the allowed regions as a result of
these variations, and in the case of debris flow, some improvement
in the quality of the fits and in the 
relative size of CoGeNT-allowed
versus excluded regions.  In the remainder of the paper, we will review
the theoretical framework (section \ref{models}) and our methods of analysis
of the various experiments (section \ref{method}).  Results are presented
in section \ref{results} and conclusions in section \ref{conc}.

\section{The models and their predictions} 
\label{models}

We briefly review the MADM model and highlight its differences 
relative to standard (inelastic) dark matter for direct detection. 
The MADM model contains a dark ``electron'' $\de$ and ``proton'' $\pd$
bound into dark atoms  $\dH$ by a new U(1) gauge force of strength
$\alpha'$, whose massless vector boson kinetically mixes with the
photon, resulting in a small electric charge $\pm \epsilon e$ for the 
$\de$, $\pd$ constituents.  MADM therefore interacts electromagnetically with protons in
ordinary nuclei.  In the special case where $m_\de = m_\pd$, the
electric coupling vanishes in the leading Born approximation
 and the dominant transition is via the
atomic hyperfine interaction. It is inelastic due to the hyperfine
splitting 
$\delta m_\dH = \frac83 \alpha'^4 m_\de^2 m_\pd^2/(m_\de+m_\pd)^3$
which can naturally be of order 10 keV.

In general, the differential
rate of DM scattering on nuclei with nuclear recoil energy $E_R$
is given by
\be 
\frac{\d R}{\d E_R}  = N_T \frac{\rho_\odot}{m_{DM}}
 \int_{v>v_{\rm min} } \d^{\,3} v\, v f(v) \frac{\d \sigma}{\d E_R} \ ,
 \ee 
where $N_T$ is the number of target nuclei, 
%$\rho_\odot \cong 0.3$
%GeV/cm$^3$ is the DM density in the solar neighborhood, 
{$\rho_\odot \cong 0.43$
GeV/cm$^3$ \cite{Salucci:2010qr} is the DM density in the solar neighborhood, }
$m_{DM}$ is the DM mass, $f(v)$ is its velocity distribution (see appendix
\ref{dmvd} for details), and $v_{\rm min}$ is the minimum DM velocity needed to
produce the given recoil energy.  In terms of the momentum transfer
$q = \sqrt{2m_N E_R}$ for nuclear mass $m_N$ and DM-nucleus reduced mass $\mu$, it is $v_{\rm min} = q/(2\mu)
+ \delta m/q$ for inelastic DM with mass splitting $\delta m$. 
The differential cross section for DM of velocity $v$ scattering on 
nuclei can be expressed as 
\be 
 \frac{\d \sigma}{\d E_R} = \frac{m_N \sigma_p}{2 v^2 \mu_n^2} 
{\left[ Z + {f_n\over f_p} (A-Z) \right]^2 } F_N^2 (q)\, 
F_{DM}^2 (q,v)
\ee
where $\sigma_p$ is the cross section for DM
scattering on protons at threshold, $\mu_n$ the DM-nucleon reduced mass,
$F_N$ the Helm nuclear form factor \cite{Lewin:1995rx}, and $F_{DM}$ the DM form factor.
For generic DM, $f_n/f_p= F_{DM}=1$, whereas for MADM, $f_n=0$ and
$F_{DM}\sim (v^2-v_{\rm min}^2)/q^2$.  

In the MADM model, the differential detection rate \jc{for
spin-independent scattering} can be 
expressed as 
\be
	{dR\over dE_R} = {\pi N_T \rho_\dH\over m_\dH\,E_R}
	\left({4\epsilon\alpha Z F_H\over
	m_\dH}\right)^2 \, I(v_{\rm min},\vec v_e)
\ee
where the velocity integral is given by
\be
	I(v_{\rm min},\vec v_e) = \int_{v_{\rm min}} {d^{\,3} v\over v} \left(v^2-v_{\rm min}^2\right)
 	f(\vec v + \vec v_e)
\ee 
\jc{(see appendix \ref{DDapp} for details).}  
Because of the $F_{DM}$ form factor, $\sigma_p$ cannot be
defined in the usual way as the cross section at threshold,
but if desired, one can define an effective $\sigma_p$
by inserting characteristic values $v_0$, $q_0$ of the DM velocity and 
momentum transfer, 
\be 
\sigma_p = \frac{64 \pi \epsilon^2 \alpha^2 \mu_n^2 v_0^2}{ m_\dH^2 
q_0^2} \ ,
\ee
for example $q_0 =25$ MeV (appropriate for scattering on 
germanium at $E_R = 5$ keV), $v_0 = 220$ km/s.  The form factor has
compensating such terms, 
\be 
F_{DM}^2 (q,v) = \frac{q_{0}^2}{q^2} 
\frac{\left(v^2-v_{\rm min}^2\right) }{v_0^2}
\label{madmff}
\ee
so that the rate is independent of $v_0$, $q_0$.

\jc{In addition, there is a spin-dependent dipole-dipole scattering 
interaction that was neglected in \cite{Cline:2012is}.  This neglect
turns out to be justified for scattering on germanium and xenon,
but not for iodine and especially sodium.  In appendix \ref{DDapp}
we give details of the dipole-dipole scattering rate, and estimates
that establish the previous statement.  Accordingly we will neglect
spin-dependent interactions in the remainder of this paper, except for
our brief study of the prospects for accommodating the DAMA annual
modulation in our global fit.}  

In the following, we will be interested not only in MADM
but also in generic DM models where
$F_{DM}=1$, which can be elastic ($v_{\rm min} = q/(2\mu)$) or
inelastic ($v_{\rm min} = q/(2\mu) + \delta m/q$).   We note that MADM
differs from the generic inelastic DM model (iDM) not only because of
the form factor (\ref{madmff}), but also because of its mild isospin
violation, $f_n=0$.  In this work we do not investigate models with
arbitrary isospin violation, such as  the case of $f_n/f_p = -0.7$
that significantly weakens the sensitivity of xenon-based experiments
\cite{Chang:2010yk,Feng:2011vu,An:2011ck,Cline:2011zr,Frandsen:2011ts,
Schwetz:2011xm,Farina:2011pw}.  However we will consider both
the isospin conserving eDM and iDM models, and the special case
in which $f_n=0$.  The latter occurs naturally when DM interactions
are mediated by a U(1) gauge boson that kinetically mixes with the
photon.

\section{Methodology}
\label{method}

In this section we describe the computation of $\chi^2$ for the CoGeNT
modulated and unmodulated data,  DAMA modulations, and the exclusion 
contours coming from XENON100, XENON10 and CDMS-Ge (CDMS-Si gives
weaker limits for the models of interest).  Some of this is standard,
following the same procedures as the relevant experimental 
collaborations themselves.  In the case of CoGeNT, we suggest an
additional background to be subtracted (over and above the surface
contamination events discussed in \cite{collar}), namely  an
unspecified modulating background that we take to be independent  of
energy, to fit away the modulation of the CoGeNT high-energy bin
1.5--3.1 keVee.  However in the end we will find that the experimental
errors on the  modulating part of the signal are too large for any of
these details to matter for the combined fit to the modulating and
unmodulated data.

\subsection{CoGeNT unmodulated signal} 

We start by fitting to the CoGeNT unmodulated signal by itself, since
this has higher statistical significance than the annual 
modulation signal. 
{We analyzed the public release data \cite{cogentdata} provided
by  the CoGeNT collaboration by reweighting  each event using the
detector efficiency $f_\text{eff}(E)$ \cite{Aalseth:2011wp}.  To
remove the known background due to cosmogenic L-shell electron capture
events in the  region of interest, we used the fitting parameters from
the K-shell peaks provided by CoGeNT \cite{cogentdata} 
and the ratio between L-shell and K-shell decays
\cite{Bahcall:1963zza}  (see e.g.\ also 
\cite{Farina:2011pw,Fox:2011px} for the parameters of the cosmogenic
backgrounds).  To relate the observed ionization energy to the 
inferred nuclear recoil energy, we take the same quenching factor
as used by CoGeNT \cite{cogentdata}, $E/$keVee $ = Q(E_R/$keVnr$)^p$, with 
$Q=0.2$ and $p=1.12$.

Subsequent to their published results,  CoGeNT reported a new estimate of the surface event
contamination near the energy threshold \cite{collar}.  We apply the
central curve of the estimated surface event correction factor on p.\
14 of ref.\ \cite{collar}  to the energy spectrum after removing the
cosmogenic background.  A constant background of  2.685 cpd/keVee, taken
as the average of the event rate between  2 keVee and 3 keVee, is
further subtracted in order to remove the events  above $E = 1.5$
keVee, whose energy dependence is approximately constant.  This is
motivated by the fact that light DM models that account for the large
excess at lower energies predict essentially no events in this higher
energy range.   For the errors on each bin, we consider the
statistical errors based on the  original events in the bin before
efficiency correction and background subtraction,  and also the
uncertainties in the parameters of the cosmogenic background.  }

\begin{figure}[t]
\hspace{-0.4cm}
\includegraphics[width=8cm,height=5cm]{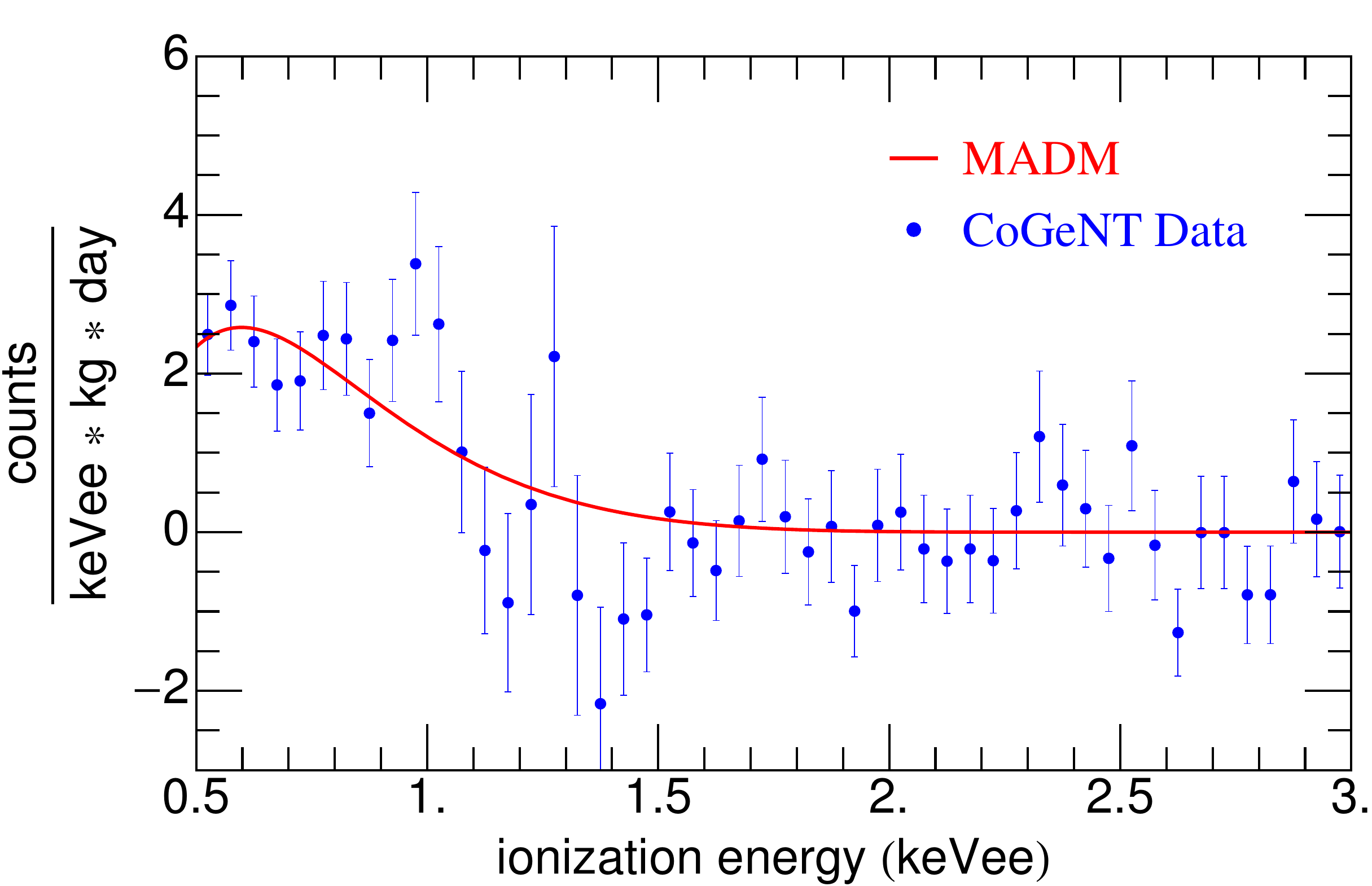}
\caption{CoGeNT energy spectrum (subtracting cosmogenic background, 
surface contamination, and constant background), and the prediction from the MADM 
model, ($m_\dH$, $\delta_\dH$, $\epsilon$)=(9.9 GeV, 24.7 keV, 0.029). 
}
\label{fig:cogent_energy}
\end{figure}

{In fig.\ \ref{fig:cogent_energy} we display the data with the cosmogenic
background,  surface events contamination, and the constant background
subtracted, along with the prediction of the MADM model with parameter
values  $m_\dH = 9.9$ GeV,  $\delta_\dH = 24.7$ keV, $\epsilon = 0.029$. 
Below we will show that this model (which we refer to as our benchmark)
corresponds to the best fit parameters and is marginally allowed by XENON
constraints, as we will determine them.}

To accurately predict the theoretical DM contribution, one should take
into account that the {442} day period of CoGeNT data-taking is not
an integer multiple of a year, as well as account for inspection and
power outage days of the CoGeNT operation, and average the modulated
rate over the actual period rather than assuming the mean value of
the  earth's velocity relative to the DM halo rest frame.  We find
that the latter makes a $-1.7$\% correction in the predicted rate, too
small  to be of interest given the current experimental uncertainties.

%%%%%%%%%%%%%%%%%%
%%%%%     figure 1
%%%%%%%%%%%%%%%%%%

%%%%%%%

\subsection{CoGeNT annual modulation}

%%%%%%%%%%%%%%%%%%
%%%%%     figure 2
%%%%%%%%%%%%%%%%%%

\begin{figure}[t]
\hspace{-0.4cm}
\includegraphics[width=8cm,height=5cm]{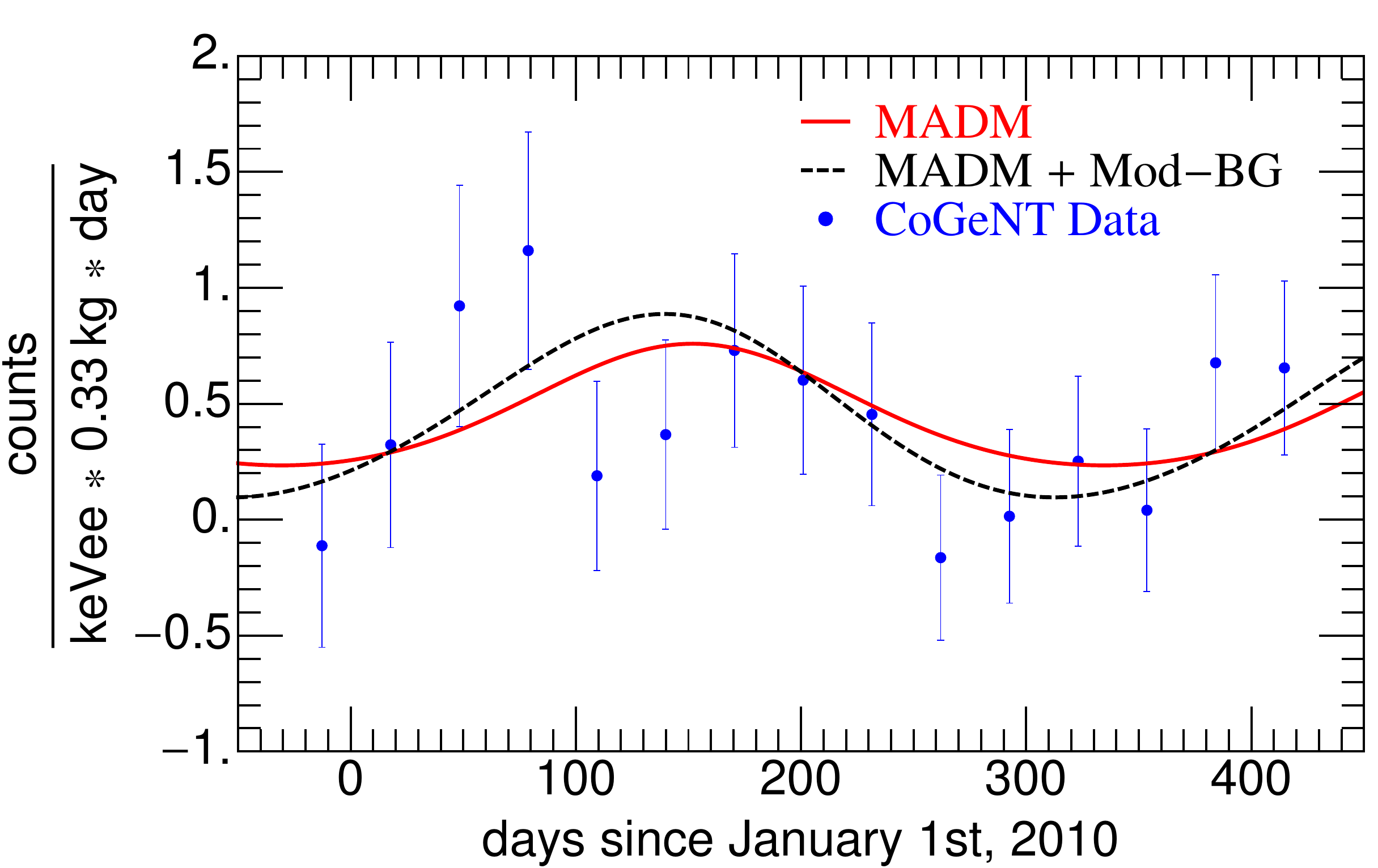}
\caption{CoGeNT data and MADM (the same model as in Fig.(\ref{fig:cogent_energy}) 
prediction for modulation in the low-energy 0.5-1.5 keVee bin.
Here we also consider the possibility of including a modulation background (dashed line) on top of the MADM signal 
in order to explain the CoGeNT modulation events in the high energy bins. The 
modulation background used here is $D \sin((2\pi/\text{year})(t-t_0))$ where 
$D=-0.487$ cpd/kg/keVee and $t_0=199$ d. As shown in the figure, 
the CoGeNT modulation amplitudes are larger than MADM predictions.}
\label{fig:cogent_lowmod}
\end{figure}
%%%%%%%

\begin{figure*}[t]
\hspace{-0.4cm}
\centerline{
\includegraphics[width=7cm]{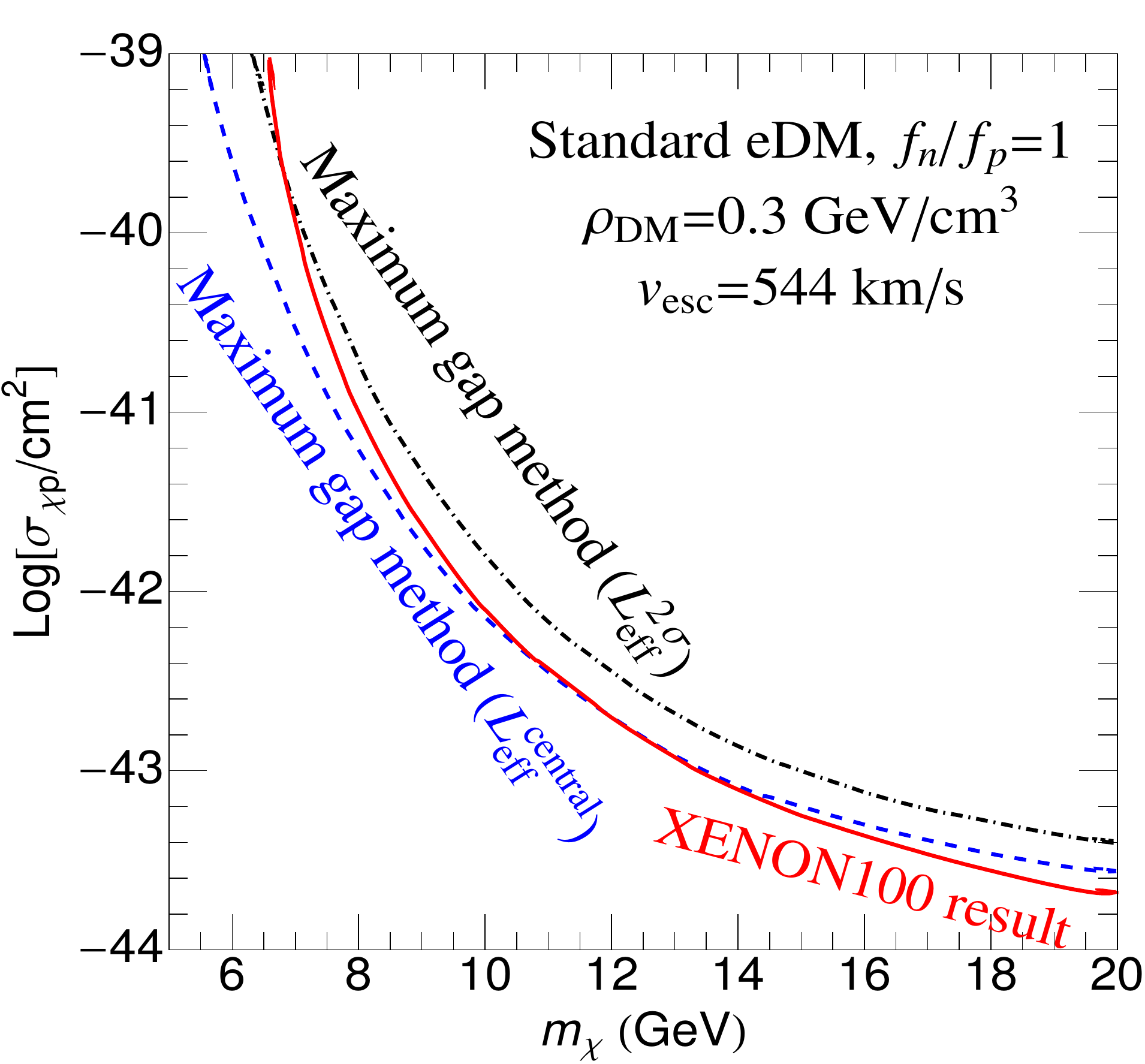}
\includegraphics[width=10cm]{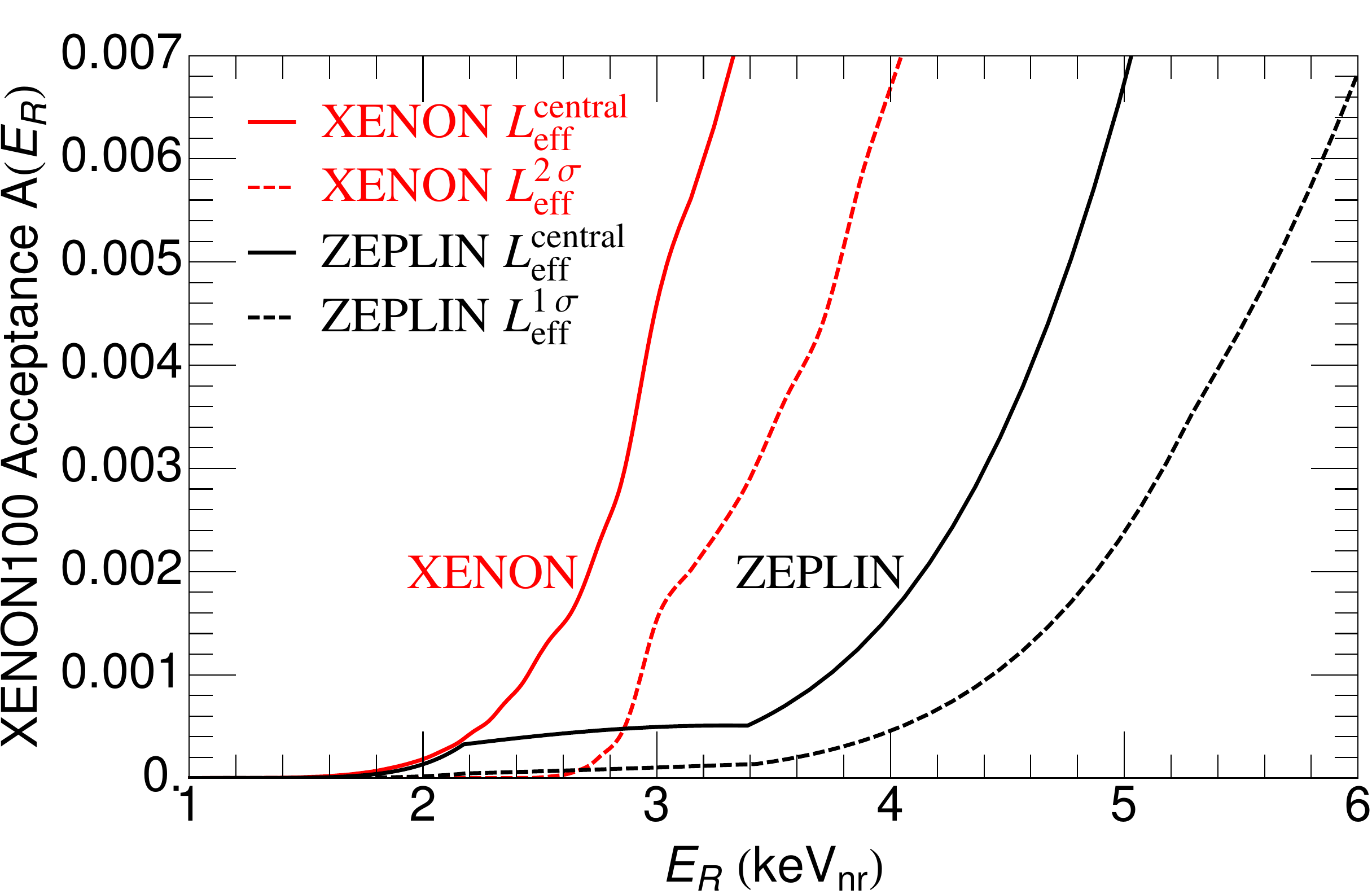}}
\caption{(a) Left: comparison of 90\% c.l.\ limits on elastic DM cross section
derived by XENON100 \cite{Aprile:2011hi} using profile likelihood method 
(solid, red curve) and by us using maximum gap method with central value 
(dashed, blue curve) and $2\sigma$
lower limit (dot-dashed, black curve) of the XENON100 relative scintillation efficiency 
${\cal L}_{\rm eff}$.
(b) Right: XENON100 acceptance curves $A(E_R)$ for the 
central values (solid lines)
and $2\sigma$ or $1\sigma$ lower limits (dashed lines) of 
${\cal L}_{\rm eff}$, taken from ref.\ \cite{Aprile:2011hi}
(left curves) and \cite{Horn:2011wz} (right curves).  The DM mass is taken
to be 10 GeV for the choice of cut acceptance.}
\label{fig:xenon_acc}
\end{figure*}
%%%%%%%

We analyzed the modulated data obtained from the CoGeNT collaboration,
following ref.\ \cite{Fox:2011px} by  grouping the events into a
low-energy bin with $E < 1.5$ keVee and a high-energy bin between
$1.5$ and $3.1$ keVee.   {To determine the modulation spectrum, we
model the cosmogenic background  by taking into account the outage
days. We also remove the constant background  and the  estimated
surface contaminated events \cite{collar} assuming that they are
unmodulated in time.  This is supported by the fact that the
surface events originally rejected by the CoGeNT analysis of
ref.\ \cite{Aalseth:2011wp} do not show any sign of modulation.  }
Based on the fit to
the unmodulated data, where no dark matter events were found to
occur above 1.5 keVee, we consider the modulating signal in the
high-energy bin to be due to an unspecified background.  To model it
using the smallest number of parameters, we take this extra
background to be independent of energy, having a rate of the form
\be
%	D \cos(\omega(t-t_0))
	{D \sin(\omega(t-t_0))}
\label{modbkd}
\ee
Assuming that $\omega = 2\pi$ y$^{-1}$ so that only $D$ and the phase
$t_0$ are free parameters,  we find that the
best fit values are {$D=-0.487$ cpd/kg/keVee and $t_0=199$ d}.
It will be seen that these values do not change very much when we allow them
vary in the minimization of the total $\chi^2$ for all the CoGeNT data.
The energy dependence of this hypothetical background is unknown, 
so we considered the two cases where it either reduces only the
signal in the high-energy bin, or that in both.   We did not see significant
differences between the two possibilities in the final results.

In any case, one needs to separate the data into an unmodulated part
(which has already been considered above) and the modulating component
whose contribution to $\chi^2$ is computed separately from that of the
unmodulated signal.  We compared several different ways of doing this:
fitting the time dependence in a given bin to a constant plus
sinusoidal term; splitting the data into its time-averaged value plus
fluctuations; or defining the time-dependent part by subtracting the
time average only over consecutive 365 day periods rather than {442}
days (which should be more correct).  However, the
data are not yet at the level of precision where these distinctions
are important.  We find that the time-averaged part of the rate in
the low-energy bin is  $2.47\pm 0.1$ cpd/keVee (corresponding to the
range $\delta \chi^2 = 1)$.

The raw CoGeNT modulation data from the low-energy bin are plotted in fig.\
\ref{fig:cogent_lowmod}, along with the MADM prediction plus modulating
background (\ref{modbkd}) for the same model parameter values as used in
fig.\ \ref{fig:cogent_energy}. Although the amplitude of the theoretical
prediction looks small  compared to the central values of the data, the
experimental error bars are large, {and so the contribution to the total
$\chi^2$ from  this discrepancy is relatively small.{ For the benchmark
model, ($m_\dH$, $\delta_\dH$, $\epsilon$)=(9.9 GeV, 24.7 keV, 0.029), 
removing the expected DM contribution from the modulation signal in the 
low-energy bin (0.5-1.5) keVee  ({\it i.e.,}\ keeping only the modulating
background  (\ref{modbkd}) as signal in this energy range),  corresponds to a
reduction in the total CoGeNT $\Delta\chi^2$ of  only $\sim 0.5$.  }

To display the regions of parameter space preferred by the data
(see section \ref{results}),
we use the $\Delta \chi^2 = 3.53,\ 8.03,\ 14.2$ contours, corresponding
to confidence intervals of 68.3\%, 95.5\%, and 
99.7\%,  appropriate for fitting three model parameters ($m_\dH$, $\delta_\dH$, $\epsilon$). 

%%%%%%%%%%%%%%%%%%
%%%%%     table  1
%%%%%%%%%%%%%%%%%%

\begin{figure*}[t]
\hspace{-0.4cm}
\includegraphics[width=18cm]{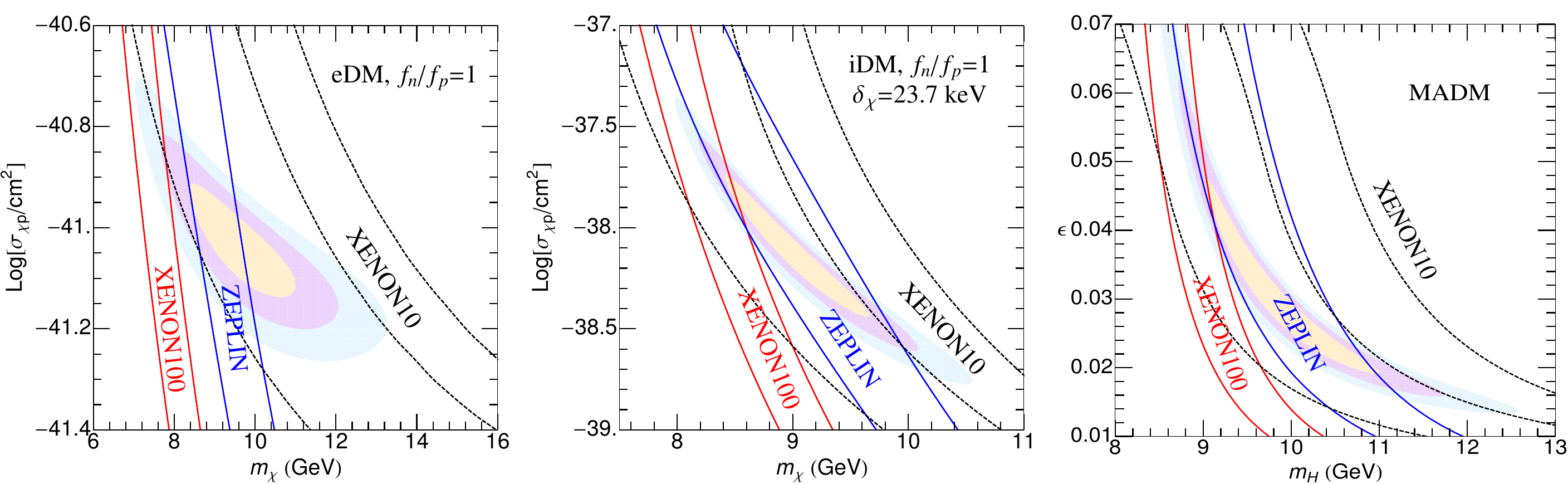}
\caption{Solid curves: limits from XENON100 assuming the
the central and 2$\sigma$ lower
limits of the XENON100 ${\cal L}_{\rm eff}$ (leftmost, red lines) and
central and 1$\sigma$ lower limit of the ZEPLIN-III ${\cal L}_{\rm eff}$
(rightmost, blue lines).  Dashed curves (black): limits from XENON10 assuming the three
possible electron yields versus energy shown in fig.\ \ref{fig:xenon_Qy}.
The shaded areas are the CoGeNT best-fit regions for eDM (left graph)
iDM (central graph) and MADM (right graph) models, discussed in
section \ref{results}.
}
\label{fig_leff_comp}
\end{figure*}
%%%%%%%

\subsection{XENON100}
\label{x100sect}

The XENON100 collaboration \cite{Aprile:2011hi} observes 3 events at
nuclear recoil energies of {12.1, 30.2 and 34.6} keV respectively, all
of which are too high to be  directly accounted for by our best-fit
models.  We compute the number of events above the XENON100
low-energy threshold  $8.4$ keVnr using Poisson fluctuations 
\cite{Felixthesis}: $N = \int dE_R \, dR/dE_R\, A(E_R)$ times the
exposure $100.9$ kg $\times\ 48$ d.  The acceptance function $A(E_R)$
depends upon  data quality cuts
and S2/S1 energy resolution  discrimination cuts which we take from
\cite{Aprile:2011hi}. For the data quality cut acceptance,  a linear
interpolation based on XENON100's analysis of the DM mass dependence
is used.  We elaborate upon this below.

Ref.\ \cite{Aprile:2011hi} uses the profile likelihood method
\cite{Aprile:2011hx} to derive its limits on dark matter.  It would be
difficult to reproduce these results without having full access to the
data (including backgrounds).  As a simpler alternative, we adopt the
maximum gap method of ref.\ \cite{Yellin2002}.  In this method,
the energy range of the search  window, 8.4--44.6 keVnr,
is  partitioned into 4 intervals (``gaps'') separated by the three
observed  events: 8.4--12.1, 12.1--30.2, 30.2--34.6, 34.6--44.6.
One then looks for the interval in which
the observation of no events is the least probable for a given DM
model, and uses this to set a limit on the cross section.  Not
surprisingly, for the low-mass DM models we are interested in, 
this turns out to be the lowest interval, 8.4--12.1 keV.

We determine the  expected number of events in each gap including not
only the dark matter signal but also the expected background, which is
estimated to be 1.8$\pm$0.6 events \cite{Aprile:2011hi} in the full
window.  We assume this background is uniformly distributed in
energy.   Since nearly all the light DM events in the full window are
within the first gap,  the DM contribution to the number of events
$x$  expected within this gap is nearly the same as the DM
contribution to the total number $\mu$ in the full search window, and the
function $C_0(x,\mu)$ that gives the Poissonian probability that more
than 0 events should have been observed in this window simplifies to 
$C_0(x,\mu) = 1 - e^{-x}(\mu-x+1)$.  The 90\% c.l.\ limit is
determined by setting $C_0 = 0.9$.  Both $x$ and $\mu$ get
contributions from DM and from the background, but the DM contribution
to each is nearly equal for light DM.  Therefore $\mu-x$ is determined
by the number of expected background events outside the first gap,
$\mu-x = 1.62$. The number of DM events allowed at the $100\,C_0$\%
confidence level  is then given by $x_{DM} = \ln(2.62/(1-C_0)) -
0.18$, where $0.18$ is the number of background events expected in the
lowest gap.  For $C_0=0.9$ this gives $x_{DM}= 3.1$.  

In fig.\ \ref{fig:xenon_acc}(a) we show that the maximum gap method gives
reasonable agreement with the more sophisticated  profile likelihood method
at masses above $8$ GeV.  At low masses, the maximum gap method gives a
stronger limit than the XENON100 result \cite{Aprile:2011hi},  unless we
weaken the assumed sensitivity by adopting the 2$\sigma$ lower limit of
XENON100's relative scintillation efficiency ${\cal L}_{\rm eff}$ (more about
which below).  This is not surprising since the profile likelihood method
essentially marginalizes over ${\cal L}_{\rm eff}$, which gives a better fit
to light DM models (weaker limit) for smaller ${\cal L}_{\rm eff}$.  In the
higher mass range,  the maximum gap method gives a limit that is close to the
one derived from XENON100's central ${\cal L}_{\rm eff}$.  We will need to
weaken this limit as described presently in order to accommodate the
CoGeNT-preferred region.  But it should be kept in mind that the additional
weakening we invoke is not so great a departure from XENON100's central
${\cal L}_{\rm eff}$ value as one might at first think, for dark matter
masses below 9 GeV.  This is because of our finding that the maximum gap
method yields stronger limits in this mass range, for a given choice of ${\cal L}_{\rm
eff}$, than does XENON100's profile likelihood approach.

To compute the expected number of DM events in a given energy
interval, one needs to weight the raw event rate $dR/dE(E_R)$ by
the acceptance function $A(E_R)$, details of whose computation are
given in appendix \ref{xenon100app}.   The acceptance at low recoil
energies depends strongly upon the relative scintillation efficiency
${\cal L}_{\rm eff}$ at low $E_R$ \cite{Savage:2010tg,Collar:2010nx}.  
The ${\cal L}_{\rm eff}(E_R)$ used
by XENON100 in their own analysis is measured down to $E=3$ keV 
\cite{Plante:2011hw} and logarithmically extrapolated to zero for lower
energies.  Previous measurements of ${\cal L}_{\rm eff}(E_R)$ had only
gone as low as  4 keV.  The validity of the new measurement at the
lowest energy has been questioned  in ref.\ \cite{Collar:2011wq},
which claims that it should be regarded only as an upper limit on the
true efficiency of the XENON100 detector.  

It will be seen that XENON100 excludes most of the parameter space
preferred for fitting the CoGeNT data unless some modest reduction of
the assumed sensitivity at low recoil energies is made.  In the 
spirit of our ``optimistic'' analysis, and taking into account the
reservations mentioned above, we adopt the recent measurement of
${\cal L}_{\rm eff}$ by the ZEPLIN-III collaboration \cite{Horn:2011wz},
which is also used in ref.\ \cite{Collar:2011wq}, and supersedes the
earlier ZEPLIN-III measurement \cite{Lebedenko:2008gb} that was criticized in ref.\ 
\cite{Savage:2010tg}.  
This determination
of  ${\cal L}_{\rm eff}$ goes down to $E_R=2.2$ keV.  
We follow the practice of ref. \cite{Savage:2010tg} in linearly extrapolating it 
from this point to zero at $E_R=1$ keV to give some  reasonable
estimate of the response at lower energies.  Using an ${\cal
L}_{\rm eff}$ close to this one, we will show that substantial overlap
between the XENON100 and CoGeNT allowed regions can be achieved.  In section
\ref{results} we will display the XENON100 constraints using the
0.5$\sigma$ lower limit (derived from \cite{Horn:2011wz}) of the
measured  ${\cal L}_{\rm eff}(E)$.  To illustrate the impact these
different assumptions make on the sensitivity of the XENON100 experiment to
low-recoil events, we plot in  fig.\ \ref{fig:xenon_acc}(b) the full
acceptance function at low recoil energies, based upon the  ${\cal
L}_{\rm eff}$ used by XENON100 \cite{Aprile:2011hi} (along with its $2\sigma$
lower limit) and that of ZEPLIN-III (showing also its $1\sigma$ lower
limit).  As we will discuss below, our  benchmark MADM model  ($m_\dH$,
$\delta_\dH$, $\epsilon$)=(9.9 GeV, 24.7 keV, 0.029)  predicts only
2.8 events using  the ZEPLIN-III 0.5$\sigma$ lower limit 
$\mathcal{L}_\text{eff}$,  {which is less than the XENON100 90\% c.l.\
upper limit of 3.1 events.}  

To show the effect of the different possibilities for  ${\cal
L}_{\rm eff}$ on the allowed DM parameters, we display in figure \ref{fig_leff_comp} the resulting
XENON100 limits on the eDM, iDM and MADM models for the central and
2$\sigma$ lower limits of the XENON100 ${\cal L}_{\rm eff}$ (solid red
curves), as well as for the central and 1$\sigma$ lower limit of the
ZEPLIN-III ${\cal L}_{\rm eff}$ (solid blue curves). The ZEPLIN-III $1\sigma$
error bars are comparable in size to the $2\sigma$ range of XENON100
since the latter is based upon an average over several different
experiments.  For reference, the CoGeNT-allowed regions for eDM, iDM
and MADM models are also shown (these will be discussed in section
\ref{results}).  Although the CoGeNT regions are ruled out by the
XENON100 central and $2\sigma$ ${\cal L}_{\rm eff}$s, and  and
marginally by the ZEPLIN-III central ${\cal L}_{\rm eff}$, one sees that
there is large overlap with the ZEPLIN-III $1\sigma$ ${\cal L}_{\rm eff}$,
and even with the ZEPLIN-III $0.5\sigma$ ${\cal L}_{\rm eff}$.  To be
conservative in our optimism, we will adopt the ZEPLIN-III $0.5\sigma$ ${\cal L}_{\rm
eff}$ for the remainder of our analysis, but it should be kept in mind
that one could reasonably argue for a case that makes even greater
room for  a viable DM explanation for the CoGeNT events.

\begin{figure}[t]
\hspace{-0.4cm}
\includegraphics[width=8cm,height=5cm]{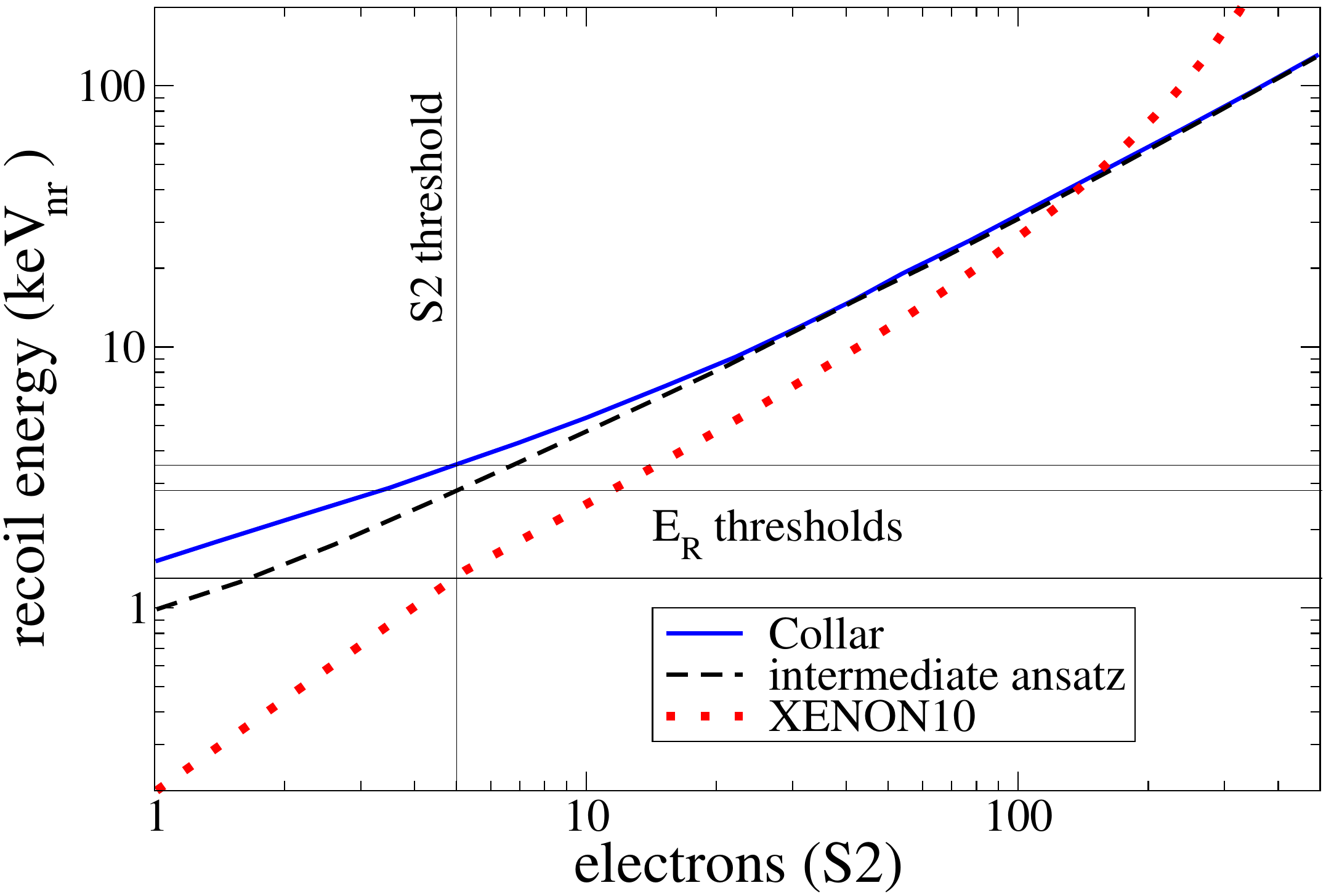}
\caption{Different possibilities for the relation between nuclear
recoil energy and number of S2 electrons observed by XENON10.
Top curve (solid) is from ref.\ \cite{Collar:2010ht,Collar:2011wq}, 
bottom curve (dotted) is that used by
\cite{Angle:2011th}.
The intermediate (dashed) curve is our hypothetical ansatz that allows
for compatibility with the CoGeNT events.}
\label{fig:xenon_Qy}
\end{figure}
%%%%%%%

\subsection{XENON10}

 Unlike XENON100, the XENON10 experiment relinquishes its demand for
scintillation photoelectrons (S1) to veto electron recoils, in order to
gain a lower energy threshold and thus greater sensitivity to light
dark matter.  Thus one does not require a significant 
acceptance function, and need
only consider the raw rate of nuclear recoils in  energy
intervals bounded by events within the range $S2 = (5,43)$.
This corresponds to recoil energies $E_R$ in the range
$(E_{\rm min},\, E_{\rm max})$ where $E_{\rm
min}$ is associated with the minimum
number of 5 electrons imposed by the experiment as a cut on the
data, and $E_{\rm max}\cong 10$ keV, whose exact value is not
important for constraining light DM. XENON10 observes 23 events in
this range of the S2 signal.   

To set a limit, energies are assigned to each event according to the
relation $S2 = E_R{\cal Q}_y(E_R)$, where  ${\cal Q}_y(E_R)$ is the
electron yield function that must be  estimated from a theoretical
model.  Then expected numbers of events $x_i$ in every energy interval
bounded by two events are compared to measured events $N_i$ in these
intervals (excluding the contributions from the endpoints),  allowing
for the computation of a probability $p_i$ that more events should
have been seen than measured, based upon Poisson statistics.  Finding
the maximum of $p_i$ over all intervals and comparing to the known
function $\bar p_{\rm Max}(C,\mu)$  leads to excluded regions in the
parameter space at a specified confidence level $C$, where $\mu$ is
the expected number of events in the interval $(E_{\rm min},\, E_{\rm
max})$.  This is the $p_{\rm Max}$ method of ref.\  \cite{Yellin2002},
which was used by XENON10 to set limits on light dark matter in
\cite{Angle:2011th}, and  in the analysis of ref.\
\cite{Farina:2011pw}.  We too use this method to compute the
90\% c.l.\ limits from XENON10. 

The crucial question is what the actual form of  the electron response
function ${\cal Q}_y$ should be, and the resulting  threshold energy
$E_{\rm min}$.  XENON10 relies upon an extrapolation of Lindhard's
theoretical prediction below 4 keV, which leads to  the estimate
$E_{\rm min}=1.4$ keV. This has been criticized by refs.\
\cite{Collar:2010ht,Collar:2011wq}, which advocate a different
prediction that leads to lower estimated sensitivity of the experiment
to light DM events, and the higher threshold $E_{\rm min}=3.6$ keV. 
We illustrate the differences between the two estimates in fig.\ 
\ref{fig:xenon_Qy}.  The XENON10 version of ${\cal Q}_y$ leads to 
a strong conflict with the regions of parameter space preferred
for explaining the CoGeNT signal, while the relaxed assumption removes
this tension entirely.  We will consider a hypothetical intermediate
ansatz, also shown in fig.\ \ref{fig:xenon_Qy}, that falls between 
the two estimates.
In fig.\ \ref{fig_leff_comp} we show the XENON10 limits that result
from the three choices of ${\cal Q}_y$ corresponding to fig.\ 
\ref{fig:xenon_Qy}; these are the dashed (black) curves.  Comparing
to the CoGeNT-allowed regions, one sees that it is sufficient to
choose the hypothetical intermediate ${\cal Q}_y$ ansatz for
maintaining compatibility between CoGeNT and XENON10.  We adopt
this choice in what follows.

\subsection{CDMS} 
For CDMS-Ge, we follow the procedure of ref.\
 \cite{Farina:2011pw}
in assigning a $\chi^2$ of the form 
\be
	\chi^2 = {(N-36)^2\over 36}\Theta(N-36)
\label{cdms_chi2}
\ee
where $N$ is the number of DM events in the relevant energy
range, from the low-energy threshold of 2 keV up to 20 keV, 
using the efficiency for the best-performing detector T1Z5 as provided by the
experiment.  The ability of CDMS to
distinguish 2 keV recoils from background has been disputed in ref.\ 
\cite{Collar:2011kf}, which argues that 5 keV could be considered as a
realistic threshold considering the uncertainties.  However we are
not forced to consider such a weakening of the ensuing CDMS
constraints since even with 2 keV threshold, it will be seen that 
they do not conflict with the CoGeNT preferred regions.
The previously observed tension between CoGeNT and CDMS-Ge has been 
alleviated by the subtraction of the CoGeNT surface contamination 
events, which has reduced the former ``signal'' events by  
$\sim$ 70\% near the low energy threshold.  In addition to including
(\ref{cdms_chi2}) in our total $\chi^2$ (although it makes no
contribution in the relevant regions), we plot the 90\% upper limit
from CDMS-Ge corresponding to $45.9$ events.  We have
also analyzed constraints from CDMS-Si data and found them to be
weaker, and therefore do not consider them henceforth.

Recently CDMS has also placed limits on a modulating DM signal in the
energy range   $5$-$12$ keVnr \cite{Ahmed:2012vq}.  However, considering
the CoGeNT quenching factor, this essentially coincides with the
high-energy bin in which our preferred models predict no events. 
Therefore we are not constrained by these data, and do not consider
them in our analysis.

We note that an independent analysis of CDMS data by ref.\ 
\cite{Collar:2012ed} finds an excess of low-energy events that is consistent
with the CoGeNT signal.  This finding is also consistent with our own observation
that the CDMS constraint is the weakest of the null results, and that we did
not have to make any special assumptions to achieve compatibility with
the CoGeNT allowed regions that we determine below.

\subsection{DAMA} 

We have computed the effect of DAMA's annual modulation data on our
fits, following the same procedure as in ref.\ \cite{Farina:2011pw}. 
\zc{As shown in appendix \ref{DDapp}, the dipole-dipole interaction
between dark matter and the target nuclei (I, Na) in the DAMA
experiment can be larger than the spin-independent
interaction, especially in collisions with large inelasticity.
Here we compute DAMA signals including contributions from
both dipole-dipole  and
dipole-charge interactions. The minimum value of $\chi^2\cong {68}$
without DAMA increases to {147} when DAMA data are included, even
though only 8 data points (the number of energy bins) are included. }
Unlike the CoGeNT allowed region, which can be made compatible with
even the most stringent conflicting data by making the judicious 
choices discussed above, the DAMA allowed regions are firmly ruled
out.   For this reason we do not try to reconcile both CoGeNT and DAMA
results with  the Xenon constraints.   We conclude that the DAMA
signal is due to some unknown background in the context of our models,
as other recent studies also seem to indicate
\cite{HerreroGarcia:2012fu}.

\begin{table}[t]
%\vspace{1cm}
    \begin{center}
\begin{tabular}{|l|c|c|c|}
\hline\hline
Maxwellian halo   & eDM & iDM & MADM \\\hline
\hline
parameters &  \multicolumn{3}{|c|}{Best-fit model} \\\hline
\hline 
$m_\chi$ or $m_\dH$ (GeV)  & 9.71 & 8.99  & 9.86\\\hline
$\delta_\chi$ or $\delta_\dH$ (keV)  & - & 23.7 & 24.7 \\\hline
$\sigma_{\chi p}$(cm$^2$) or $\epsilon$ & $10^{-40.3}$ &
$10^{-37.4}$ & 0.029 \\\hline
$-D$ (cpd/kg/keVee)  & 0.55  & 0.50 & 0.49 \\\hline
$t_0$ (day) & 205 & 200 & 198  \\\hline
\hline
CoGeNT & \multicolumn{3}{|c|}{$\chi^2$} \\\hline
\hline
energy spectrum  & 53.9 & 45.4 & 46.4\\\hline
modulation (0.5-1.5) keVee  & 7.60 & 8.58 & 8.94\\\hline
modulation (1.5-3.1) keVee  & 9.36 & 9.40 & 9.47\\\hline
total      & 70.9 & 63.4 & 64.8\\\hline
\hline
\end{tabular}
\caption{
The best-fit models for CoGeNT in standard eDM, iDM, and in
MADM models.
All the models in the table have isospin violation $f_n=0$. 
The case of $f_n=f_p$ for eDM or iDM is the same except for decreasing $\sigma_{\chi p}$
by a factor of $(A/Z)^2 = 5.15 = 10^{0.71}$.
}
 \label{tab:best}
\end{center}
 \end{table}

\begin{figure}[t]
\hspace{-0.4cm}
\includegraphics[width=0.5\textwidth]{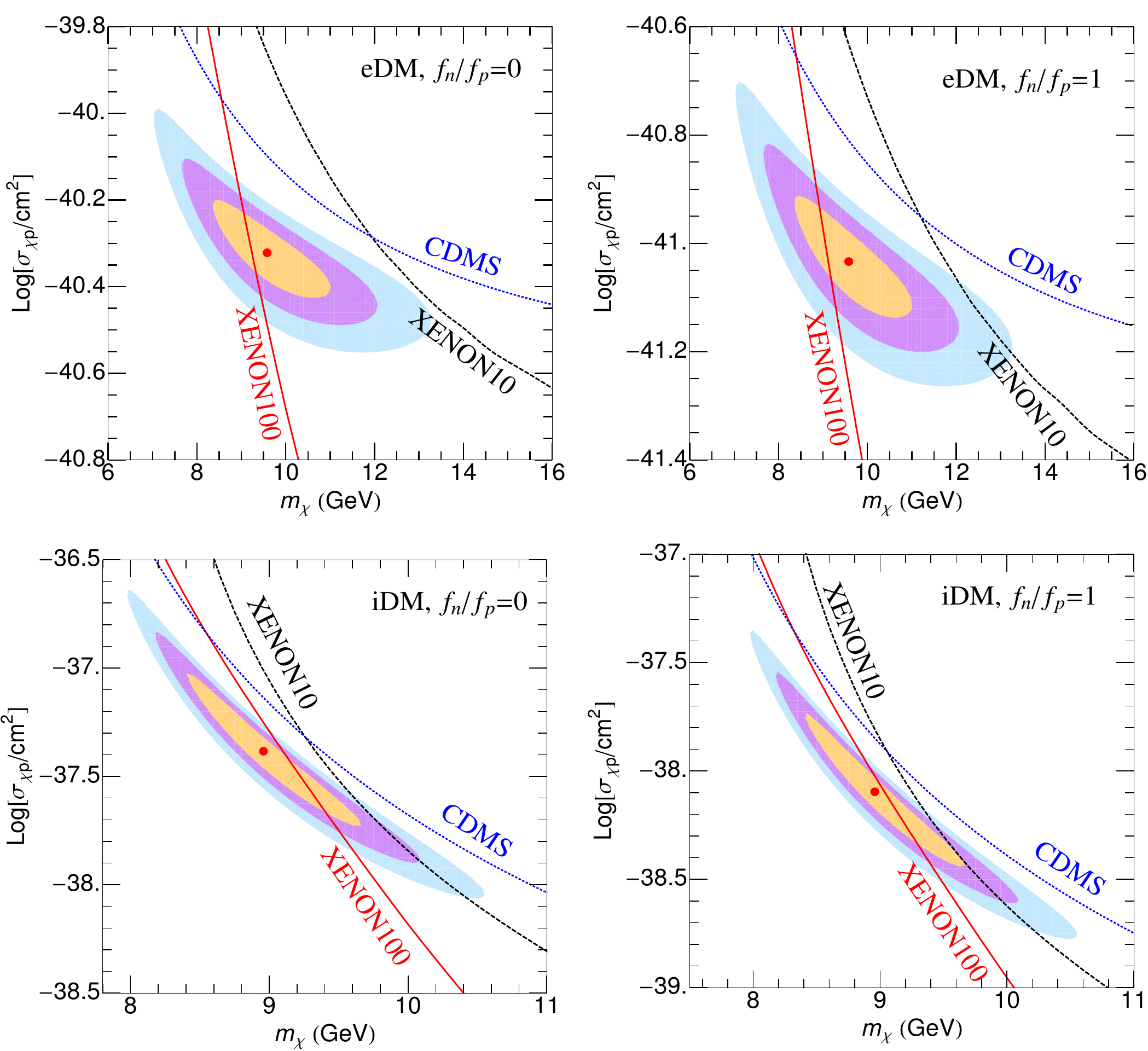}
\caption{CoGeNT best-fit regions (shaded) in the
$m_\chi$-$\sigma_{\chi p}$ plane, 
 and ``optimistic'' 90\% c.l.\ upper limits from XENON10,
XENON100 and CDMS-Ge for eDM (top row) and iDM (bottom row) models,
with isospin violation $f_n=0$ (leftmost graphs) or isospin conservation
(rightmost graphs).  The shaded regions correspond to 68.3\%, 95.5\% and
99.7\% confidence intervals.}
\label{fig:panel-vanilla}
\end{figure}

\begin{figure*}[t]
\hspace{-0.4cm}
\includegraphics[width=\textwidth]{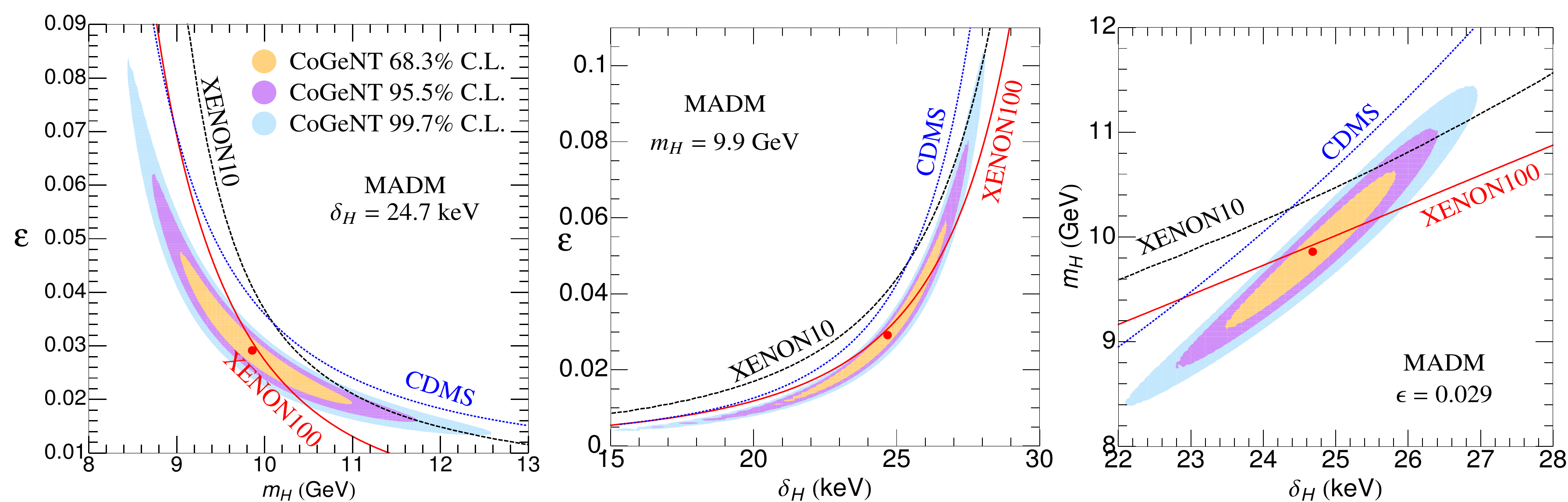}
\caption{CoGeNT best-fit regions (shaded) for the MADM model
near the benchmark point ($m_\dH$, $\delta_\dH$,
$\epsilon$)=(9.9 GeV, 24.7 keV, 0.029).
The shaded regions correspond to 68.3\%, 95.5\% and
99.7\% confidence intervals.  In each plane, the parameter that is not
varying is fixed to its best-fit value.}
\label{fig:panel-madm}
\end{figure*}

\section{Results}
\label{results}

\subsection{Standard DM velocity distribution}

We determined the regions of parameter space that give the best fits
to the three models under consideration: standard elastic and
inelastic dark matter (denoted as eDM and iDM, respectively), and the
millicharged atomic (MADM) model.  For eDM, there are just two model
parameters, the DM mass $m_\chi$ and cross section on nucleons $\sigma_n$,
while iDM also includes the mass splitting $\delta_\chi$. For MADM, we
use the fractional charge $\epsilon$ to parametrize the interaction strength
instead of $\sigma_n$, and we distinguish the mass and mass splitting
from those of standard DM using  $m_\dH$ and $\delta m_\dH$ (where
$\dH$ stands for the dark  ``hydrogen'' atom).    For all models, we
adopt the relaxed versions of the XENON100 and XENON10 constraints
corresponding to the ZEPLIN-III $0.5$-$\sigma$ lower limit on ${\cal
L}_{\rm eff}$ (fig.\  \ref{fig_leff_comp}) and the hypothetical
intermediate ${\cal Q}_y$ (fig.\ \ref{fig:xenon_Qy}), respectively.

In table \ref{tab:best} we display the parameters corresponding to the
models that minimize the $\chi^2$ for the CoGeNT data (as well as CDMS
although the latter gives a vanishing contribution to $\chi^2$ at these
best fit points), as well as the corresponding minimum values of
$\chi^2$.   Inelastic DM with $m_\chi\cong 9$ GeV and $\delta_\chi\sim 24$ keV gives a
significantly better fit than does elastic, and the millicharged
atomic dark matter model, also inelastic, gives nearly as good a fit
as iDM, with a slightly larger mass $m_\dH \cong 10$ GeV and 
mass splitting $\delta_\dH\sim 25$ keV.  
Due to the relatively large
inelasticity, the DM-nucleon cross section in  the iDM case
is about 3 orders of magnitude higher than that of eDM.  

For the eDM and iDM models, fig.\ \ref{fig:panel-vanilla} displays the
CoGeNT 68.3\%, 95.5\% and 99.7\% confidence intervals, along with the
constraints from  XENON10, XENON100 and CDMS, in the plane of $m_\chi$
and the cross section for DM scattering on protons.  The XENON100
limits are the most constraining, cutting through the center of the
CoGeNT allowed regions, but still leaving room for compatibility
within the 68.3\% c.l.\ intervals.  The best-fit points are excluded
by XENON100 for eDM but allowed for iDM.  Having untuned isospin
violation of the form $f_n=0$ (DM coupling only to protons) slightly weakens
the Xenon constraints relative to the CoGeNT regions.

Next we turn to the MADM model, which we study in somewhat greater
detail. The best-fit parameter set for this model is marginally consistent with
XENON100, and we adopt it as our benchmark model.
In table \ref{tab:benchmark} we present
the contributions for this model to $\chi^2$ from the
nonmodulated and modulated data and from comparison to the
DAMA annual modulation, and numbers of events expected for XENON and
CDMS experiments.  
In figure \ref{fig:panel-madm}, the best-fit regions and 90\% c.l.\
constraints for MADM are displayed in the three planes
$m_\dH$-$\epsilon$, $\delta_\dH$-$\epsilon$, and $\delta_\dH$-$m_\dH$,
where in each case the parameter that is not varied is fixed to its
benchmark value.  Approximately half of the CoGeNT 68\% confidence
intervals are left unexcluded by the XENON100 limit, which is the
most stringent of the constraints.

\begin{table}[tbh]
%\vspace{1cm}
    \begin{center}
\begin{tabular}{|l|c|c|c|}
\hline\hline
Experiment &   Data & MADM  & $\chi^2$ \\\hline
\hline
CoGeNT energy spectrum & - & - & 46.5 \\\hline
CoGeNT modulation (0.5-1.5) keVee & - & - & \po 8.7 \\\hline
CoGeNT modulation (1.5-3.1) keVee & - & - & \po 9.4 \\\hline
CoGeNT total    & - & - & 64.7 \\\hline
\hline
CDMS-Ge (T1Z5), (2-20) keVnr & 36  & 26.1 & 0 \\\hline
\hline 
XENON100,  $L_\text{eff}^\text{central}$  & - & 32.1  & -  \\\hline
XENON100,  $L_\text{eff}^{1 \sigma}$  & - &  18.3 &  - \\\hline
XENON100,  $L_\text{eff}^{2 \sigma}$  & - &  10.0 &  - \\\hline
XENON100,  ZEPLIN $L_\text{eff}^\text{central}$  & - & \po 5.7  & -  \\\hline
XENON100,  ZEPLIN $L_\text{eff}^{0.5 \sigma}$  & - &  \po 2.8 &  - \\\hline
XENON100,  ZEPLIN $L_\text{eff}^{1 \sigma}$  & - &  \po 1.1 &  - \\\hline
\hline 
XENON10, their ${\cal Q}_y$ & 23  &  49.0 &  -  \\\hline
XENON10, intermediate ${\cal Q}_y$ & 23  & \po 5.8 &  -  \\\hline
XENON10, Collar ${\cal Q}_y$ & 23  & \po 1.2 &  -  \\\hline
\hline 
%DAMA modulation & -  & - & 84.4 \\\hline
DAMA modulation & -  & - & \zc{79} \\\hline

\hline
\end{tabular}
\caption{
Contributions to the total $\chi^2$ or to the number of events
observed by CDMS-Ge or XENON experiments, for the benchmark
MADM model with
($m_\dH$, $\delta_\dH$, $\epsilon$)=(9.9 GeV, 24.7 keV, 0.029).  
For the CoGeNT modulation spectrum, we include the modulating
background (\ref{modbkd}).
For the XENON100 and XENON10 limits, we consider various choices
for ${\cal L}_{\rm eff}$ and ${\cal Q}_y$, discussed in section
\ref{method}.
}
 \label{tab:benchmark}
\end{center}
 \end{table}

\subsection{Nonstandard DM halos}

We have  investigated the effects of two possible modifications
to the standard Maxwellian phase space density of the dark matter. The
first is a distortion of the shape, of the type $f(v^2;v_{\rm
esc}^2)\to f^k(v^2/k; v_{\rm esc}^2/k)$ (see eq.\ (\ref{fk})) that has
been suggested in ref.\ \cite{Lisanti:2010qx} based upon Eddington's
relation between $f(v)$ and and the density profile $\rho(r)$.
Comparisons
to $N$-body simulations and to the Milky Way halo give 
$k = [1.5,3.5]$ and $k\sim 2$ respectively.  The effects of varying
$k$ on
fitting to CoGeNT were previously studied in ref. \cite{Fox:2011px}.
We find that increasing
$k$ has a relatively small effect on the best-fit regions and on the
constraint curves, shifting them to somewhat higher values of DM mass
and cross section, and smaller values of the mass splitting.  We
illustrate this for the MADM model  in figure \ref{fig:k-comp},
where the best-fit values of the parameters and the minimum value of
$\chi^2$ are plotted as functions of $k$.

\begin{figure}[t]
\hspace{-0.4cm}
\includegraphics[width=0.5\textwidth]{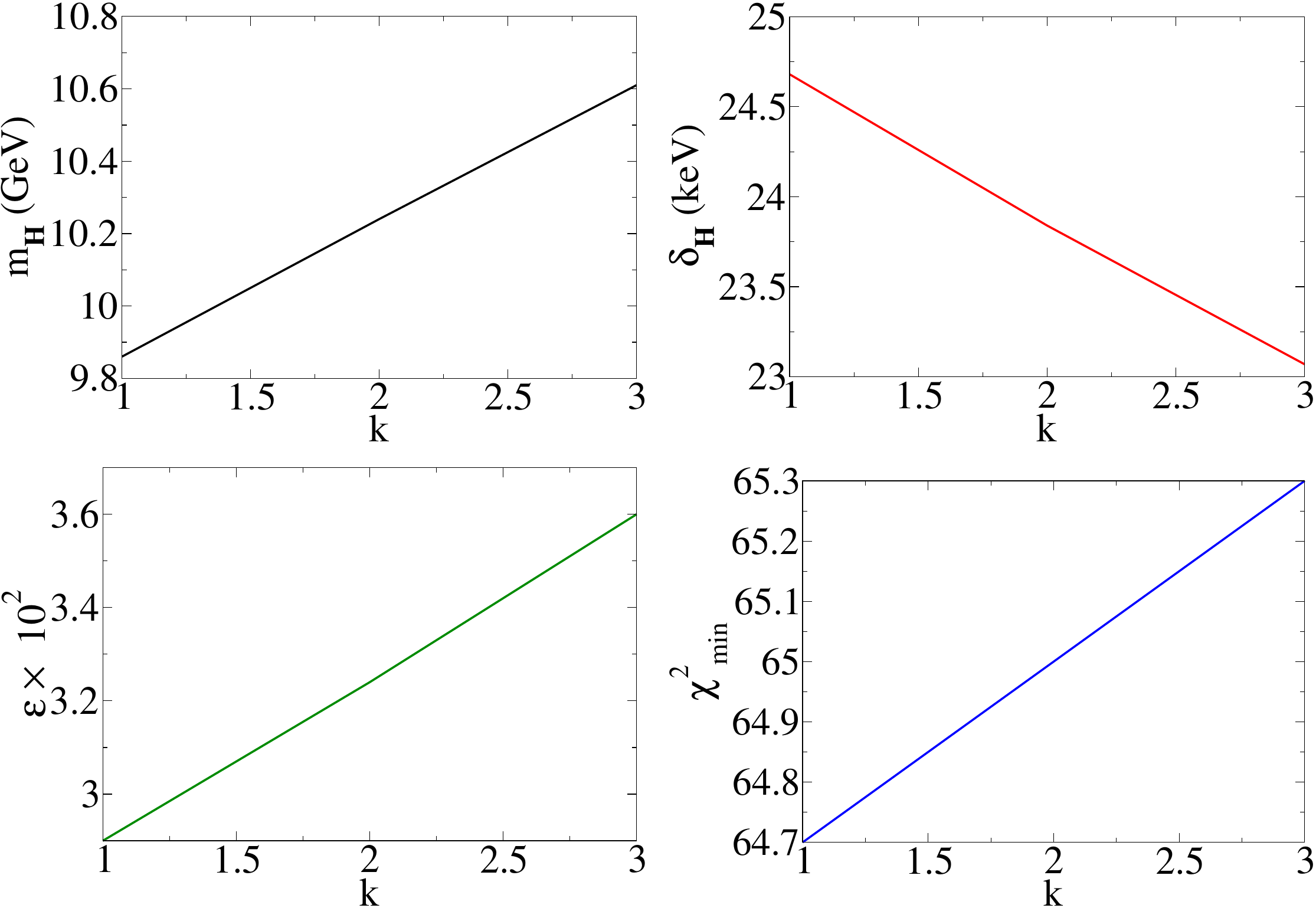}
\caption{Dependence of best fit values of $m_\dH$, $\delta_\dH$,
$\epsilon$ and minimum value of $\chi^2$ on the halo velocity profile
index $k$, in the MADM model.}
\label{fig:k-comp}
\end{figure}

We next consider the debris flow scenario of ref.\
\cite{Kuhlen:2012fz}.  Tidal debris is a form of unvirialized dark
matter due to tidal stripping of ubiquitous subhalos in the galaxy. 
Unlike tidal streams, which are rare features whose effects have been
previously considered for interpreting CoGeNT in ref.\
\cite{Natarajan:2011gz}, debris flow is argued in \cite{Kuhlen:2012fz}
to be homogeneously distributed in the galaxy and therefore guaranteed
to be present in the solar system, with a distribution that was
inferred in \cite{Kuhlen:2012fz} by analysis of $N$-body simulation
results.  

We correct an error \cite{kuhlen} in \cite{Kuhlen:2012fz} by writing
the full distribution function as 
\be
	f(v) = (1-\epsilon_0)\, f_{v}(v) + \epsilon_0\, f_{d}(v)
\label{full_f}
\ee
where $f_{v}$ is the Maxwellian distribution of the virialized
component, $f_d$ is the contribution from debris, and $\epsilon_0 =
0.22$ is determined by comparison to $N$-body simulations.  
In \cite{Kuhlen:2012fz}, the coefficients of $f_{v,d}$ were mistaken for functions
of velocity in the computation of the DM detection rate.  Ref.\ 
\cite{Kuhlen:2012fz} further models $f_{d}(v)$ (in the earth frame)
by an isotropic
function with support in the window $v_f-v_e < v < v_f + v_e$,
\be
	f_{d}(v) \cong {v\over 2 v_f\, v_e}
	\Theta(v_f+v_e-v)\Theta(v - v_f + v_e)
\label{debris_f}
\ee
where $v_e(t)$ is the modulating earth velocity and $v_f$ is a flow
velocity characterizing the debris.  ($f_{v,d}$ are both
normalized such that $\int dv f_{v,d} = 1$.)  In \cite{Kuhlen:2012fz},
the value $v_f = 340$ km/s was suggested, but we find that a higher
value $v_f = 475$ km/s is required for consistency with their
determination of the ratio $\epsilon(v_{\rm min})$
of debris particles to the total number of DM particles
having $v>v_{\rm min}$.  In fig.\ \ref{fig:lisanti-comp} we show that
$\epsilon(v_{\rm min})$ as directly computed from eqs.\ 
(\ref{full_f},\ref{debris_f}) is in reasonable agreement with the result from
\cite{Kuhlen:2012fz} based on analysis of $N$-body simulation data,
for our choice of $v_f$, which is definitely not the case for 
$v_f = 340$ km/s.
This provides a consistency check on the approximation
(\ref{debris_f}) with $v_f=475$ km/s.  
The constant $\epsilon_0$ in (\ref{full_f}) is equal to 
$\epsilon(0)$.

\begin{figure}[t]
\hspace{-0.4cm}
\includegraphics[width=0.4\textwidth]{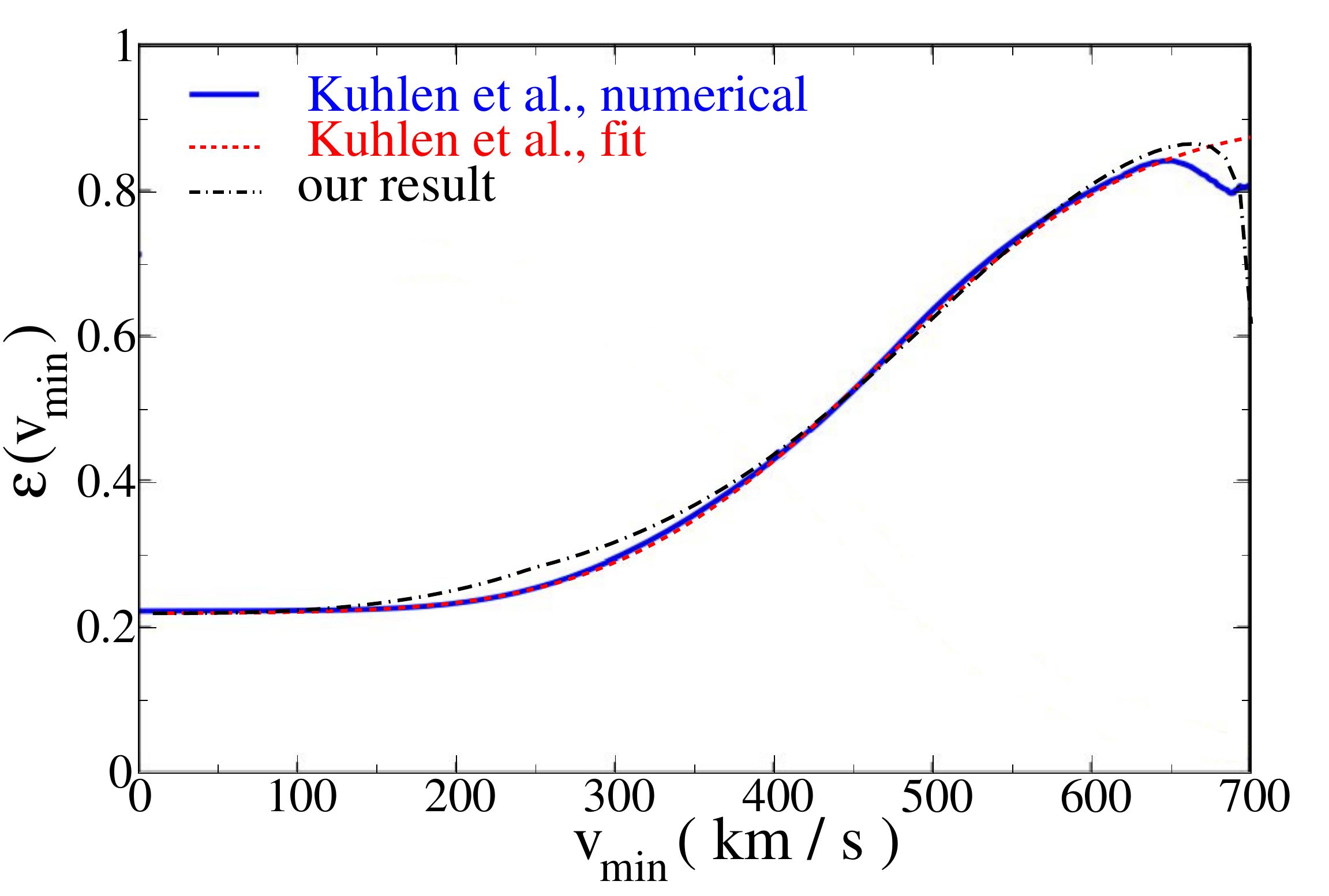}
\caption{The ratio of debris particles to total number of dark matter
particles having $v>v_{\rm min}$ in the earth frame, as determined
in ref.\ \cite{Kuhlen:2012fz} (solid and dot-dashed curves), and as
calculated here, using the approximation (\ref{debris_f})
 in eq.\ (\ref{full_f}). }
\label{fig:lisanti-comp}
\end{figure}

\begin{table}[t]
%\vspace{1cm}
    \begin{center}
\begin{tabular}{|l|c|c|c|}
\hline\hline
Debris flow   & eDM & iDM & MADM \\\hline
\hline
parameters &  \multicolumn{3}{|c|}{Best-fit model} \\\hline
\hline 
$m_\chi$ or $m_\dH$ (GeV)  & 7.4 & 9.8 (9.0)  & 10.9 (9.8)\\\hline
$\delta_\chi$ or $\delta_\dH$ (keV)  & - & 20.7 (23.6) & 22.6 (24.8) \\\hline
$\sigma_{\chi p}$(cm$^2$) or $100\,\epsilon$ & $10^{-40.9}$ &
$10^{-39.8}$ ($10^{-38.0}$) & 0.63 (3.6) \\\hline
$-D$ (cpd/kg/keVee)  & 0.55  & $0.51$ ($0.50$) & $0.49$ ($0.49$) \\\hline
$t_0$ (day) & 205 & 202 (200) & 199 (199)  \\\hline
\hline
$\chi^2$ & 71.1 & 59.8 (63.3) &  63.9 (64.7)\\\hline
\hline
\end{tabular}
\caption{
The best-fit models for CoGeNT in standard eDM and iDM and
MADM models, including the debris flow component in the DM
phase space distribution function.
The eDM and iDM models are  isospin conserving, $f_n=f_p$.
Numbers in parentheses refer to the second-best fits in case
of a second local minimum of $\chi^2$.
}
 \label{tab:debris}
\end{center}
 \end{table}

\begin{figure*}[bth] \hspace{-0.4cm}
\includegraphics[width=\textwidth]{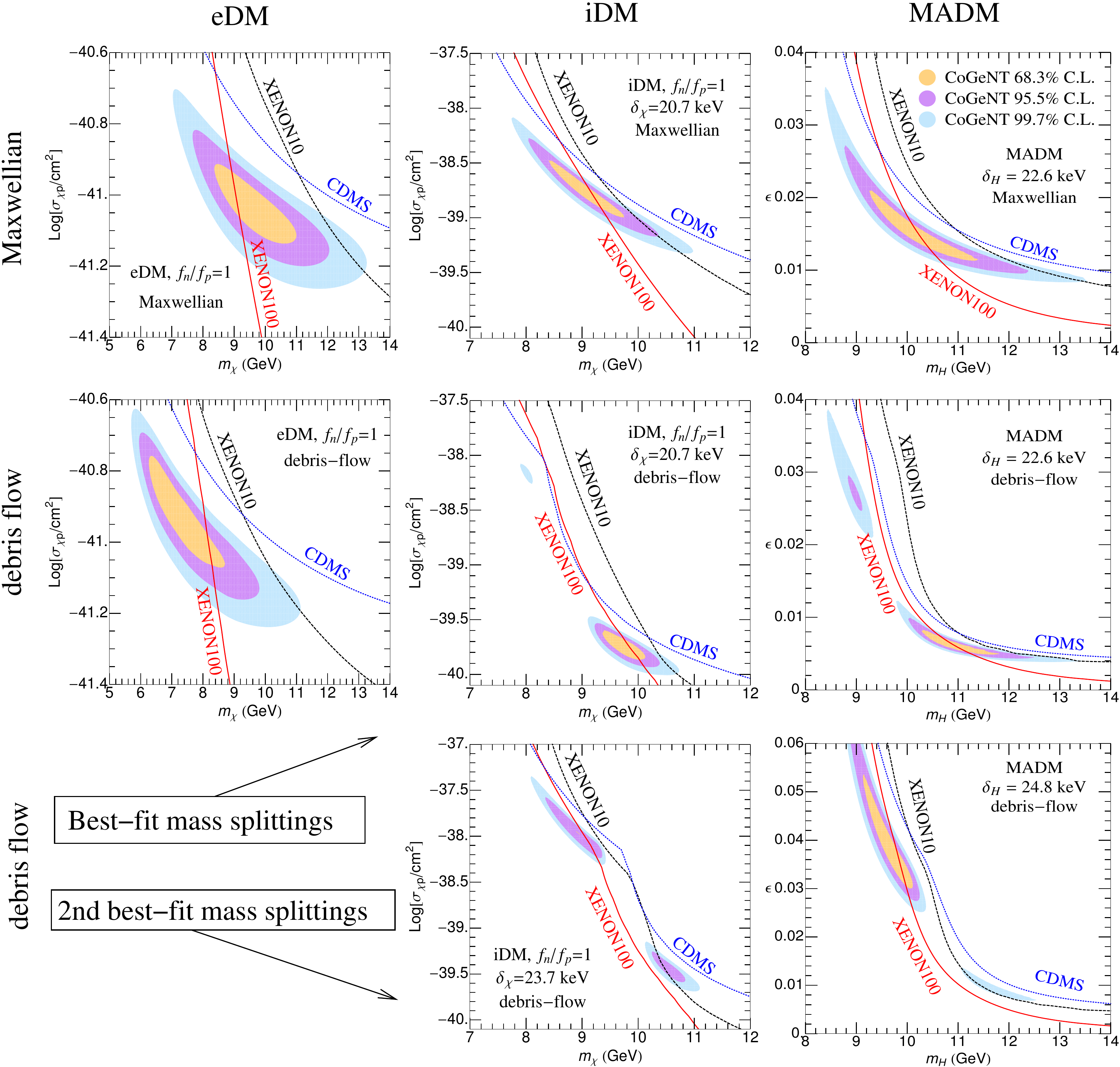}
\caption{CoGeNT best fit regions and constraints for the eDM, iDM and
MADM models (left to right) for Maxwellian DM velocity distribution
(top row) and including   debris flow contribution (second and third
rows).  The middle row sets the mass splittings to the values
corresponding to the global minima of the $\chi^2$ for iDM and MADM,
while the bottom row  corresponds to the  higher local minima.   The
eDM and iDM results are for the isospin-conserving $f_n=f_p$ case.}
\label{fig:debris} \end{figure*}

The debris flow contribution has interesting effects on the quality
of fits and the best-fit regions for dark matter direct detection.  
We summarize the best-fit values in table
\ref{tab:debris}.
The graphical results are presented in 
fig.\ \ref{fig:debris}.  In the case of isospin-conserving 
elastic DM, the goodness of
fit is nearly unchanged, but the best-fit mass is shifted to the
lower value $m_H = 7.4$ GeV, while the cross section
increases\footnote{Recall that one must divide the best-fit $\sigma_{\chi,p}$
values in table \ref{tab:best} by 5.15 to convert them to the
isospin-conserving case} by
a factor of $\sim 5$, thus moving below the region excluded by
XENON100, whose constraints are also shifted, but by a lesser amount.

For inelastic DM, debris flow leads to an improvement in the best fit
with $\chi^2 = 59.8$ (smaller by 3.6 than the Maxwellian halo
result).  The most dramatic change amongst the best-fit model
parameters relative to the Maxwellian halo values is in the lower
value of the cross section.  There is no noticeable relative shift
between the CoGeNT best-fit region and the  Xenon constraints in this
case.  However a new local minimum with $\chi^2 = 63.3$ develops at
lower $m_\chi$, higher $\delta_\chi$  and larger $\sigma_{\chi p}$. 
It originates from the presence of two components in $f$, only one of
which dominates the signal near a given local minimum. {The new
low-mass region requires higher recoil velocity for a given recoil
energy (especially at large mass splittings), and the debris flow
contribution dominates at high $v$.}   A hint of this second minimum is
visible in the lower middle graph of fig.\ \ref{fig:debris}, which has
$\delta_\chi = 20.7$ keV, the value corresponding to the global
minimum, rather than the $\delta_\chi$ of the secondary minimum.  It
is clearly visible in the bottom middle graph where $\delta_\chi \cong
24$ keV. An interesting feature of the  second minimum is that, even
though its $\chi^2$ is higher by 3.5  than at the global minimum, it
evades the Xenon constraints somewhat more robustly.

The MADM model with debris flow shares qualitative features
similar to those of iDM.  The best-fit $\chi^2$ is better
(though only by 0.8) than in the Maxwellian case, the DM
mass is higher, and the fractional charge $\epsilon$ is smaller.
There is still a second local minimum, but in this case its
mass splitting  and $\chi^2$ are closer to that of the global minimum, so that
both regions appear more clearly on the same contour plot at the
global $\delta_\dH$ best-fit value.  Again, the secondary minimum
at lower mass and higher cross section is less constrained by
XENON.

\zc{
\subsection{Update on XENON100 Analysis}

After the first version of this paper, the XENON100 collaboration 
announced their new limits on DM-nucleus scattering
\cite{Aprile:2012nq}. In this subsection, we give an indication
of how this changes our previous analysis by reconsidering only the
MADM model.  Here we also 
consider the $1\sigma$ lower limit of the ZEPLIN-III  
${\cal L}_{\rm eff}$ rather than just $0.5\sigma$ as used above.  
Employing the maximum gap method (see Appendix \ref{xenon100-2012}
for details), we derive the limits from the new XENON100 data shown 
as dashed lines in fig.\ \ref{fig:xe2012-madm}, corresponding to the $0.5\sigma$ and $1\sigma$  
limits on ${\cal L}_{\rm eff}$, respectively.  This figure should be
compared to 
the leftmost graph in Fig.(\ref{fig:panel-madm}).  Somewhat
surprisingly, 
the new XENON100 $90\%$ c.l.\ exclusion curve crosses the
previous XENON100 curve in the region of MADM parameter space shown
here.   The reason that they do not remain parallel at low DM mass
is that events below 3 keVnr are not used in the new
XENON100 analysis for exclusion limits
\cite{Aprile:2012nq}.  Comparing the two figures, we 
see that the newly excluded region on
MADM based on the recent XENON100 data \cite{Aprile:2012nq} is
nearly the same as the old one 
under the same ${\cal L}_{\rm eff}$ assumption. 
%With a
%more conservative ${\cal L}_{\rm eff}$ choice,  the allowed region
%remains almost the same as our previous XENON100 analysis. 

\begin{figure}[htbp]
\hspace{-0.4cm}
\includegraphics[width=0.4\textwidth]{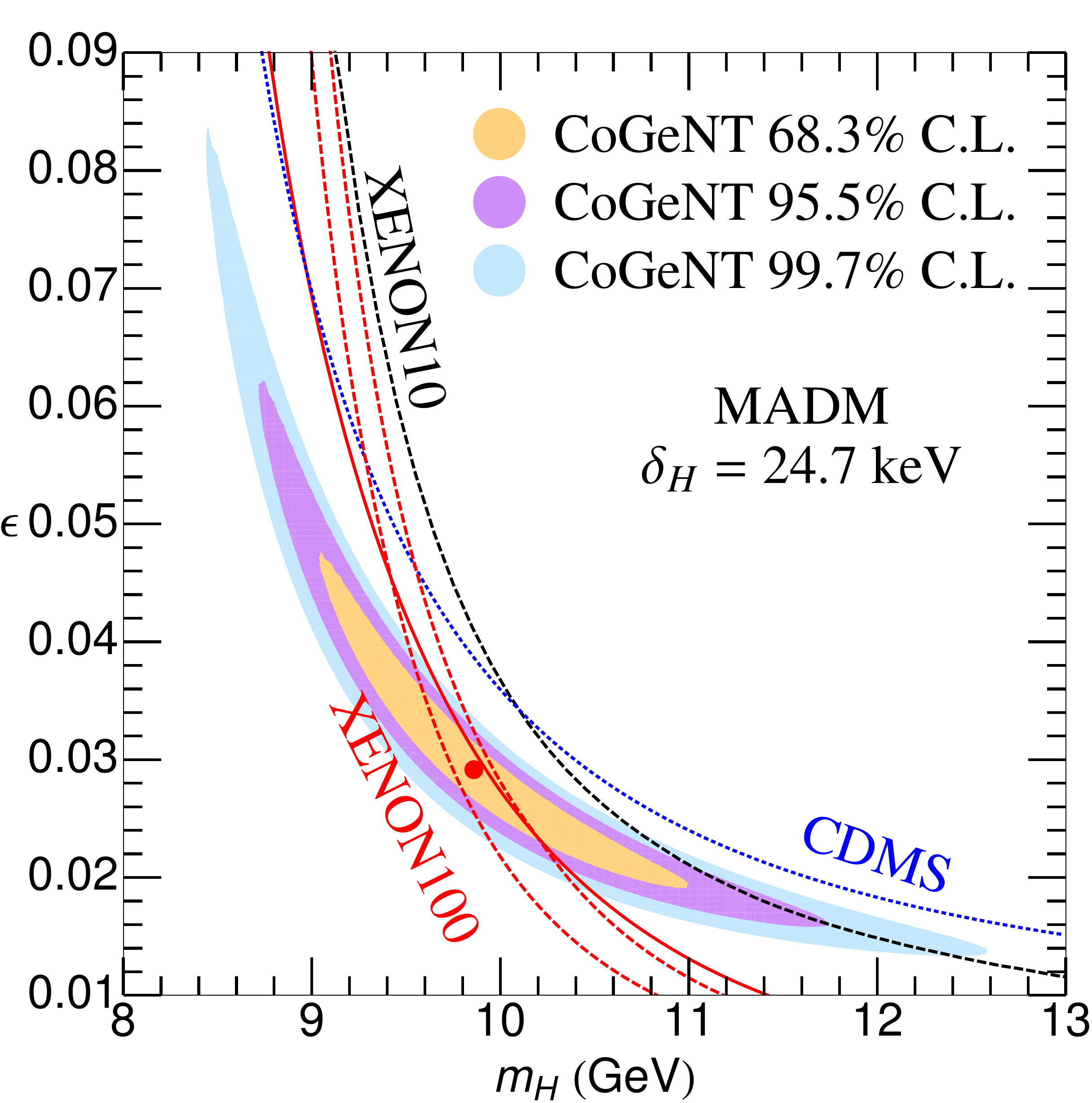}
\caption{\zc{CoGeNT best-fit regions (shaded) for the MADM model
$\delta_\dH$= 24.7 keV. The two red dashed lines are the new 
XENON100 limits \cite{Aprile:2012nq} assuming ZEPLIN-III 
$0.5\sigma$ and $1\sigma$ lower limits on ${\cal L}_{\rm eff}$.}}
\label{fig:xe2012-madm}
\end{figure}

}

\section{Conclusions}
\label{conc}

In this work we have attempted to find the most credible set of
assumptions that would allow for a dark matter interpretation of the
CoGeNT excess, without any fine-tuning of isospin-violating couplings,
while remaining at least marginally consistent with constraints from
null experimental results, mainly those of XENON100 and XENON10.  Our focus was
on standard elastic and inelastic DM (possibly having mild isospin
violation $f_n=0$), and on a particular example of magnetic inelastic
DM, the millicharged atomic DM model.  Some elements of our analysis
that are new relative to previous studies are: subtraction of 
recently identified
surface-contaminated CoGeNT events; hypothesis of a modulating
background, especially at high recoil energies; and consideration of
an unvirialized DM debris flow component (correcting an error in the
literature).  

The main arguments for believing that the DM interpretation of CoGeNT
can still be viable are the experimental or theoretical uncertainties
in the Xenon detector response sensitivities (parametrized by the 
functions ${\cal L}_{\rm eff}$ and ${\cal Q}_y$ respectively, for
XENON100 and XENON10) at low recoil energies.  For XENON10 to exclude
the CoGeNT preferred region, a theoretical model for ${\cal Q}_y$ is
required \cite{{Angle:2011th}}, which is argued to be no more
plausible than an alternate model \cite{Collar:2010ht,Collar:2011wq}
that leaves room for the CoGeNT allowed region.  We showed that an
ansatz for ${\cal Q}_y$ intermediate between these two extremes is
also adequate for sufficiently softening the XENON10 constraint.
Similarly XENON100's ability to exclude the CoGeNT region relies upon
a particular measurement and extrapolation to low energies of the
relative scintillation efficiency ${\cal L}_{\rm eff}$.  We showed that the assumption of an
alternate ${\cal L}_{\rm eff}$, well within the error bars of another
recent measurement \cite{Horn:2011wz}, is sufficient to make the
CoGeNT best-fit model marginally compatible with the XENON100
constraint.

While standard elastically scattering DM can still give a reasonable
fit to all the data, inelastic models do better.  Our featured MADM
model has a different dependence upon the recoil momentum and DM
velocity than standard iDM, but fits the data nearly as well as
standard iDM.   Our assumption of a hypothetical modulating background
to explain CoGeNT modulations above $1.5$ keVee improves the $\chi^2$s
somewhat, but our results depend quite weakly upon this feature,
due to the lower statistical significance of the modulations compared
to the unmodulated signal.  Distortions from the Maxwellian DM
distribution functions of the form (\ref{fk}) have a small effect on
our results.  However an additional contribution from debris flows 
(including a correction original to this paper) noticeably lowers the
DM mass and relaxes the tension between eDM and XENON100,  improves
the fit of iDM, and has the interesting effect of introducing a second
local minimum in the $\chi^2$ of iDM and MADM at lower DM mass, whose
tension with Xenon is reduced relative to the global best-fit model.

Although our results for standard eDM and iDM models may be of primary
interest for most readers, our original motivation was to study the
validity, as an explanation for CoGeNT, of MADM, which appeared to be
a new class of model; only in the course of this work was its
similarity to other inelastic magnetic DM models noticed.  But the
existence of constituents with fractional electric charge $\epsilon
\sim 0.01$ is a distinguishing feature of MADM, and a potentially
worrisome one since early studies of similar models claimed that 
$\epsilon \gtrsim 10^{-3}$ is ruled out \cite{Goldberg:1986nk} by searches
for anomalously heavy isotopes of hydrogen.  In ref.\ 
\cite{Cline:2012is} we pointed out several possible loopholes to
this constraint, including the fact that the screening effect of
the earth's magnetosphere on the subdominant ionized component of the
dark matter has not been taken into account in its derivation. 
Subsequent to the publication of \cite{Cline:2012is}, we have received
confirmation from some of the authors on the heavy isotope search
papers \cite{klein_etal,hemmick_etal} that the sensitivity of these
searches to isotopes with charges deviating from exact integer values
could have been significantly reduced \cite{hkp}.  We therefore
keep an open mind about the viability of MADM even with fractional
charges as large as $\sim 0.01$ as we find necessary for explaining
CoGeNT.  It could be interesting to reinvestigate in more detail what
values of $\epsilon$ are consistent with existing searches for
charged relics.

Based on these results, we remain optimistic that CoGeNT may really be
detecting dark matter of mass $\sim 10$ GeV, and perhaps with
inelastic couplings between states split by $\sim 20$ keV.  New data
expected from CoGeNT in the near future should shed further light on 
this exciting possibility.

\section*{Acknowledgments}
We benefitted greatly from correspondence or discussions with
E.\ Aprile, 
J.\ Collar, T.\ Hemmick, F.\ Kahlhoefer, J.\ Klein, M.\ Kuhlen, M.\ Lisanti, P.\ Sorensen, 
Z.\ Whittamore, I.\ Yavin and S.\ Yellin.  Our research was supported by the
Natural Sciences and Engineering Research Council (NSERC, Canada).

\appendix

\section{\lowercase{Dark matter velocity distributions}}
\label{dmvd}

The velocity distribution function for the standard Maxwellian
halo is often taken to be 
\be
	f = N\left( e^{-v^2/v_0^2} - e^{-v_{\rm esc}^2/v_0^2}\right)
\ee
with
\be
	N^{-1} = 4\pi v_0^3\left({\sqrt{\pi}\over 4}{\rm erf}(a) -
ae^{-a^2}\left(\frac12 +\frac13 a^2\right)\right)
\ee
where $a = v_{\rm esc}/v_0$.  Standard values are $v_0 = 220$ km/s,
$v_{\rm esc} = 500-600$ km/s, $v_e = 232\pm30$ where $\pm 15.3$ is the
range of annual variation.  (The earth's velocity relative to the
DM barycenter appears in $f(\vec v + \vec v_e)$ when computing
nuclear recoil rates.)  Unless otherwise noted, we adopt 
$v_{\rm esc} = 600$ in computing detection rates.

In addition to this standard form, we also consider a generalized
expression with the extra parameter $k$,
\be
	f_k = N_k\left( e^{-v^2/k v_0^2} - 
	e^{-v_{\rm esc}^2/k v_0^2}\right)^k
\label{fk}
\ee
with normalization factor
\be
	N_k^{-1} = 4\pi\int_0^{v_{\rm esc}} dv\, v^2\,
\left( e^{-v^2/k v_0^2} - e^{-v_{\rm esc}^2/k v_0^2}\right)^k
\ee
which has no closed-form expression except for integral values of
$k$.

\section{Direct detection rate}
\label{DDapp}

In this appendix we give details of the derivation of differential
scattering on nuclei for MADM through spin-independent (dipole-charge) and
spin-dependent (dipole-dipole) interactions.  The differential
detection rate for either case is given by
\be
	{dR\over dE_R} = {2\pi N_T m_N \rho_\dH\over m_\dH\,\mu^2}
	\int_{v_{\rm min}} {d^{\,3}v\over v'} f(v) {d\sigma_N\over d\Omega}
\ee
where $v' = \sqrt{v^2-y}$ is the
final velocity, with $y\equiv 2\delta_\dH/\mu$.  
The average matrix element squared for dipole-charge
scattering is given by  \cite{Cline:2012is} 
\be
	(C/q)^2 |\vec v\times \hat q|^2 (\mu/p)
\ee
where $C = Z\epsilon e^2/m_{\de}$. 
The differential cross section for the dipole-charge interaction is given by
\bea
	{d\sigma_N\over d\Omega} &=& \left({4\epsilon\alpha Z\over
m_\dH}\right)^2 {\mu^4 v v'^3\over q^4}\sin^2\theta\, F_H^2\nonumber\\
	&=& \left({4\epsilon\alpha Z\over
m_\dH}\right)^2 \sqrt{1-{y\over v^2}}\, 
	\left(v^2-v_{\rm min}^2\right)\,{F_H^2\over x}
\eea
where we defined $x\equiv q^2/\mu^2$ and used the relations
\be
	\sin^2\theta = {x(v^2-v_{\rm min}^2)\over v^2 (v^2-y)},\quad
	v_{\rm min} = {x+y\over 2\sqrt{x}}
\label{angle}
\ee
Then the rate due to dipole-charge interaction can be written as
%\be
%	{dR\over dE_R} = {\pi N_T \rho_\dH\over m_\dH\,E_R}
%	\left({4\epsilon\alpha Z F_H\over
%	m_\dH}\right)^2 \, I(x,y,\vec v_e)
%\ee
\begin{equation}
\frac{dR_\text{DZ}}{dE_R} = \frac{N_T \rho_\dH \pi}{m_\dH}
 \left( 4 \epsilon \alpha \over m_\dH \right)^2  \frac{Z^2F_H^2}{E_R}
\left\langle \frac{v^2 - v_\text{min}^2}{v} \right\rangle 
\label{dz}
\end{equation}
where the velocity integral is given by
%\be
%	I(x,y,\vec v_e) = \int_{v_{\rm min}} {d^{\,3} v\over v} \left(v^2-v_{\rm min}^2\right)
% 	f(\vec v + \vec v_e)
%\ee
\be
	\left\langle \frac{v^2 - v_\text{min}^2}{v} \right\rangle 
	= \int_{v_{\rm min}} {d^{\,3} v} \left(v^2-v_{\rm min}^2 \over v \right)
 	f(\vec v + \vec v_e)
\ee
with $\vec v_e$ the earth's velocity relative to the halo.

%
%\section{Comparison to dipole-dipole scattering}

For the dipole-dipole interaction, we consider the Hamiltonian 
\be
	H_{\rm int} = {\epsilon e\over 3 m_{\de}}(\vec \sigma_{\de}- \vec \sigma_{\pd})
	\cdot \vec\mu_N\,\delta^{(3)}(\vec r_\dH-\vec r_N)
\ee
where the $\sigma$'s are Pauli matrices corresponding to the spins of
$\pd$ and $\de$, and $\vec\mu_N$ is the magnetic moment of the
nucleus.  The spin-summed squared matrix element is 
\be
	\sum|\langle f|H_{\rm int}|i\rangle|^2 = \frac49
	\left({I+1\over I}\right)\left(\epsilon e \mu_N\over m_{\de}
	\right)^2 (\mu/p)
\ee
%The values of $\mu_N$ can be found at 
%http://www.webelements.com/; we give them in table \ref{mutable}.
%For comparison, the average matrix element squared for dipole-charge
%scattering is given by
%\be
%	(C/q)^2 |\vec v\times \hat q|^2 (\mu/p)
%\label{dz}
%\ee
The differential detection rate due to dipole-dipole scattering is 
\begin{equation}
\begin{split}
\frac{dR_\text{DD}}{dE_R} & =  \frac{N_T \rho_\dH \pi}{m_\dH} 
\left( 4 \epsilon \alpha \over m_\dH \right)^2  
\left\langle \frac{1}{v} \right\rangle
%\\
%& \times 
\frac{2m_N F_M^2}{9m_p^2} 
\sum_i f_i \hat\mu_i^2 \frac{I_i+1}{I_i} 
\end{split}
\label{dd}
\end{equation}
where the summation is over possible isotopes of the target nucleus, 
$f_i$ is the relative abundance, $\hat\mu_i$ is the magnetic moment
of isotope $i$ in units of $e/2m_p$, and $I_i$ is the spin. 
$F_M$ is the form factor for the nuclear spin.  In the following 
estimates we will assume that $F_M\cong F_H$.  
The ratio of (\ref{dd}) to (\ref{dz}) is
\be
	R = \sum_i R_i\equiv \left( q F_M\over 3 m_p Z  F_H\right)^2 
	F(v_\text{min})
\sum_i f_i \hat\mu_i^2 \frac{I_i+1}{I_i} 
\ee
where the function $F$ is given by 
\be
F(v_\text{min}) \equiv 
\left\langle \frac{1}{v} \right\rangle /
\left\langle \frac{v^2 - v_\text{min}^2}{v} \right\rangle
\ee
When $v_\text{min}<$ 500  km/s, $F(v_\text{min})/10^6 \sim$(1-1.4). 
For large $v_\text{min}$ values, $F(600, 700, 750, 800)/10^{6}$ = (1.6, 2.2, 3.1, 7.2). 
Taking $q=20$ MeV, $\delta_\dH = 25$ keV, $m_\dH = 10$ GeV, 
we find $R_i \sim (7.1/Z)^2 (F/10^6) f_i \hat\mu_i^2 (I_i+1)/I_i $. 
We list these values in the last colummn of table \ref{mutable}. 
%For comparison, we also include the ratio when $\delta_\dH=0$ in the 
%last column. 

\begin{table}[t]
\vspace{5mm}
% \begin{minipage}[b]{1.8\linewidth}
% \centering
\caption{Relevant isotopes, atomic number, abundances, spins, magnetic moments (in
units of $e/2m_p$), minimum velocity for $(q,m_\dH,\delta_\dH)$=(20 MeV, 10 GeV, 25 keV), 
function $F(v_\text{min})$, ratio of dipole-dipole to dipole-charge scattering. 
%ratio when $\delta_\dH=0$. 
The values of $\mu_N$ can be found at \cite{magneticdipole}.}
%\begin{ruledtabular}
\begin{tabular}{|c|c|c|c|c|c|c|c|c| }
\hline
isotope & $Z$ &   \% & spin & $\hat\mu_N$ &  $v_\text{min}$ 
& $F/10^6$ & $R_i$  \\
\hline
$^{23}$Na & 11 &100 & $3/2$ & $2.218$  & 815 & 13.2 & 45  \\ 
$^{73}$Ge & 32 & 7.7 & $9/2$ & $-0.879$  & 719 & 2.5 & 0.009 \\
$^{127}$I & 53 & 100 & $5/2$ & $2.813$  & 700 & 2.2 & 0.4  \\
$^{129}$Xe & 54 & 26.4 & $1/2$ & $-0.778$ & 700 & 2.2 &  0.02 \\
$^{131}$Xe & 54 & 21.2 & $3/2$ & $0.692$ & 700 & 2.2 &  0.006 \\
\hline
\end{tabular}
\label{mutable}
% \end{minipage}
%\end{ruledtabular}
\end{table}

We have checked these estimates more quantitatively
by computing the energy spectra due to dipole-dipole (DD) and 
dipole-charge (DZ) scattering
in the  MADM model
($m_\dH, \delta_\dH, \epsilon$)=(9.9 GeV, 24.7 keV, 0.029).  In the
CoGeNT
and CDMS experiments, the DD interaction contributes about  1\% 
of the total rate.
In the XENON100 experiment, the signal due to  DD is
about 5\% of the total. For collisions on iodine in DAMA, 
the DD spectrum has roughly the same magnitude as DZ, 
but most of  the
recoil events have energies below the region of interest.  The DD
spectrum is about 20 times larger than that of DZ for collisions 
on Na in the DAMA experiment.  The recoil spectra on Na due to DD and DZ
were also compared for  the elastic case ($\delta_\dH=0$), in which
case the two contributions are of comparable size.
We conclude that
the DD interaction is important for the DAMA analysis,  
especially for
scenarios with large inelasticity.

\section{XENON100 acceptance}
\label{xenon100app}

In XENON100, the relative scintillation efficiency
${\cal L}_{\rm eff}$ is important for determining the
mean number of S1 photoelectrons corresponding to a given 
nuclear recoil energy, through the relation 
\be 
\label{S1er}
\mu_{S1}(E_R ) = \left( \frac{S_{nr} }{S_{ee}}   
\right)\,\mathcal{L}_\text{eff}( E_R )\, L_y\, E_{R} \ .
\ee
where $S_{ee} = 0.58$ and $ S_{nr} =0.95 $~\cite{Aprile:2006kx} are
the suppression in the scintillation yield for electronic and nuclear
recoil respectively, and the light yield $L_y =  2.20$ PE/keV$_{\rm
ee}$ is normalized at 122\ keV$_{\rm ee}$ \cite{Aprile:2010um}.
The uncertainties in these parameters are negligible compared to
that of ${\cal L}_{\rm eff}$.  

In order to use the experiment to constrain DM models,
one could follow the likelihood method introduced in
\cite{Aprile:2011hx}.  A simpler
method (similar to the one we adopt) is described  in the thesis \cite{Felixthesis}, which
requires the acceptance function $A(E_R)$, that enters in the number of
expected DM events in a given energy range $(E_1,E_2)$ as 
\be 
N = {\cal E}\int_{E_1}^{E_2} \frac{\d R}{\d E_R} A(E_R)\, \d E_R 
\ee
where ${\cal E} = 4843.2$ kg$\cdot$d is the exposure.  The acceptance function  
takes account of $\mathcal{L}_\text{eff}$, the Poisson
fluctuations of the $S1$ signal, and the acceptance of the $S1$ and
the nuclear recoil cut $\zeta_{\rm cut} ( n_1)$ \cite{Aprile:2011hi}, 
\be 
A( E_R) = \sum_{n_1 = S1_{\rm min}}^{S1_{\rm max}} 
P\left( n_1, \mu_{S1}(E_R) \right) \zeta_{\rm cut} ( n_1)
\ee
where $\mu_{S1}(E_R)$ is given in eq.~(\ref{S1er}) and the
S1 window used by XENON100 is $ S1_{\rm min}=4$, $S1_{\rm max}=30$.
The acceptance of the applied cuts  $\zeta_{\rm cut} $ is the product
of the data quality cut and S2/S1 electron recoil discrimination cut,
shown in fig.\ 2 of ref.\ \cite{Aprile:2011hi}.  The data quality cut
depends upon the DM mass, and we interpolate between the curves given
in \cite{Aprile:2011hi}.  $P$ is the Poisson distribution,
\vskip-0.5cm
\be 
P\left( n_1, \mu_{S1}(E_R) \right)  = 
{\mu_{S1}(E_R) ^{n_1} \e {- \mu_{S1}(E_R) }}/{n_1 !} \ .
\phantom{{{{|_|}_|}_|}_|}
\label{eq:poisson}
\ee

To compute an upper limit from the number of events, we use the
maximum gap method of ref.\ \cite{Yellin2002}, which as described
in sect.\ \ref{x100sect}  restricts $N<3.1$ at 90\% c.l.\ in the
energy window 8.4--12.1 keVnr.

%%%%%%%%%%%%%%%%%%
%%%%%     Xe 2012 
%%%%%%%%%%%%%%%%%%

\section{XENON100 analysis (2012 Data)}
\label{xenon100-2012}

Two DM candidate events with $3.3$ PE and  $3.8$ PE were observed in XENON100 data \cite{Aprile:2012nq}, 
which partition the search region S1=(3-20) PE into 3 gaps for the maximum gap analysis: (3-3.3), (3.3-3.8), and (3.8-20). The predicted DM events in each gap are computed via the acceptance function $A_i(E_R)$  
\begin{equation}
N_i =\mathcal{E} \int \frac{dR}{dE_R} (E_R) A_i(E_R) dE_R
\end{equation}
where $i=1,2,3$ and $\mathcal{E}=$ 34 kg $\times$ 224.6 days. 
The acceptance function $A_i(E_R)$ is given by  
%denoting the 3 gaps used in this analysis, 
%$A_i(E_R)$ is the acceptance function based on the 3 gaps, 
%$dR/dE$ is the differential DM rate in XENON100 detector.  
%The acceptance functions are given by  
\begin{equation}
\begin{split}
A_i(E_R) = & \epsilon_{S2} (E_R)  \sum_{N_{pe}=1}^{\infty} P(N_{pe}, \mu_{S1}(E_R)) \\
\times &
\int_{S1^i_\text{min}}^{S1^i_\text{max}} \epsilon_{S1} (S1) G(S1, N_{pe}) dS1
\end{split}
\end{equation}
where $\epsilon_{S2} (E_R)$ and $\epsilon_{S1} (S1)$ are the efficiency functions for 
$S2$ and $S1$ signals respectively which are digitized from  \cite{Aprile:2012nq}. 
 $P(N_{pe}, \mu_{S1}(E_R))$ is given in Eq.(\ref{eq:poisson}). 
 $G(S1, N_{pe})$ is a Gaussian distribution describing  the response of the PMTs  
% $S1$ is the measured value of the primary scintillation signal, and 
% $s1(E_R)$ is its expectation value. 
%(Following the convention used in \cite{Aprile:2012vw}: upper case letters 
%denote actual measured quantities, while lower case letters for expectation values.)
%\begin{equation} 
%s1(E_R) = E_R \mathcal{L}_\text{eff}(E_R,\mathcal{E}=0) 
%\mathcal{L}_y(E_\text{ref},\mathcal{E}) \frac{S_{nr}(\mathcal{E})}{S_{ee}(\mathcal{E})}
%\end{equation} 
%The Poisson fluctuations of the photoelectron is given by 
%\begin{equation}
%\text{Poisson}(N_{pe}, s1(E_R)) = \frac{{s1}^{N_{pe}} e^{-s1}}{(N_{pe})!}
%\end{equation}
%The Gaussian function is given by 
\begin{equation}
\text{G}(S1, N_{pe}) = \frac{1}{\sqrt{2\pi}} \frac{1}{\sigma_\text{PMT} \sqrt{N_{pe}}} 
\exp \left[ -\frac{1}{2}\left(\frac{S1-N_{pe}}{ \sigma_\text{PMT} \sqrt{N_{pe}}}\right)^2 \right]
\end{equation}
where $\sigma_\text{PMT}$=0.5 PE \cite{Aprile:2012vw}.  

To compute the exclusion limits in maximum gap method \cite{Yellin2002}, we use the number of DM events 
calculated here together with the expected background events, (1.0$\pm$0.2) \cite{Aprile:2012nq}.

\end{document}